\documentclass[iop]{emulateapj}

\usepackage{graphics}
\usepackage{color}
\usepackage{amsmath,amssymb}
\usepackage{type1cm}
\usepackage{ulem}
\usepackage{mathrsfs}
\slugcomment{draft version \today, }

\shorttitle{Photospheric Emission From Stratified Jets}                                                      
\shortauthors{Ito et al.}

\begin{document}

\title{Photospheric Emission From Stratified Jets}
%Resolving the $z=4.4$ Quasar Host Galaxy of BRI\,1335-0417: \\
%Intimate Interaction or Massive Molecular Disk?}

\author{Hirotaka Ito\altaffilmark{1}, Shigehiro Nagataki\altaffilmark{1}, Masaomi Ono\altaffilmark{1}, Shiu-Hang Lee\altaffilmark{1},  Jirong Mao\altaffilmark{1},  Shoichi Yamada\altaffilmark{2,3}, Asaf Pe'er\altaffilmark{4,5}, Akira Mizuta\altaffilmark{6,7}, Seiji Harikae\altaffilmark{8}}

\altaffiltext{1}{Astrophysical Big Bang Laboratory, RIKEN, Saitama 351-0198, Japan}
\altaffiltext{2}{Department of Science and Engineering,
 Waseda University, 3-4-1 Okubo, Shinjuku, Tokyo 169-8555, Japan}
\altaffiltext{3}{Advanced Research Institute for Science and Engineering, 
Waseda University, 3-4-1 Okubo, Shinjuku, Tokyo 169-8555, Japan}
\altaffiltext{4}{Institute for Theory and Computation, Harvard-Smithsonian Center for Astrophysics, 60 Garden Street, Cambridge, MA 02138, USA}
\altaffiltext{5}{Physics Department, University College Cork, Cork, Ireland}
\altaffiltext{6}{KEK Theory Center, Tsukuba 305-0801, Japan}
\altaffiltext{7}{Computational Astrophysics Laboratory, RIKEN,Saitama 351-0198, Japan}
\altaffiltext{8}{Quants Research Department, Financial Engineering Division, Mitsubishi UFJ Morgan Stanley Securities Co., Ltd., Mejirodai Bldg., 3-29-20 Mejirodai, Bunkyo-ku, Tokyo 112-8688, Japan}
\email{hito@yukawa.kyoto-u.ac.jp}

\begin{abstract}

We explore 
photospheric emissions from
stratified two-component jets, wherein a
highly relativistic spine outflow is surrounded by a wider and
less relativistic
sheath outflow.
%spine-sheath structure are explored.
%It is assumed that fast spine jet is embedded in a slower sheath outflow. 
%{\color{green}The propagation of
%thermal photons which are injected 
%in  regions of high optical depth %with local temperature 
%is solved
%until they escape at the photosphere.}
Thermal photons are injected in regions of high optical depth
and propagated until they escape at the photosphere.
Due to the presence of shear
in  velocity (Lorentz factor)
at the boundary of 
the spine and sheath region, a fraction of the injected
photons are
accelerated via a Fermi-like acceleration mechanism
such that a high energy power-law tail is formed in the resultant spectrum.
% gain energy by a Fermi-like acceleration mechanism
%and form a high energy
%power-law tail in the resultant spectrum.
%
%We show, in particular, that the result can
%successufuly reproduce the
%spectra of observed prompt emissions of gamma-ray bursts (GRBs). 
%
We show, in particular, that if a velocity shear
with a considerable variance in the bulk Lorentz factor
is present,
the high energy part of  observed Gamma-ray Bursts (GRBs) 
photon spectrum
can be explained by 
this photon acceleration mechanism. 
We also show that the accelerated photons 
may also account for the
origin of the extra hard power-law component 
above the bump of the thermal-like peak seen  in some
peculiar bursts (e.g., GRB 090510, 090902B, 090926A).
It is demonstrated that time-integrated spectra
can also reproduce the low energy spectrum of GRBs consistently
due to a multi-temperature effect when 
time evolution of the outflow is considered.
%Although the low energy part of the spectra 
%do not vary with the shear velocity,
%It is  also shown that low energy spectrum of GRBs can
%be explained consistently 
%when time evolution of the outflow is considered.
%
Finally, we show that the
empirical $E_{\rm p}$-$L_{\rm p}$
relation can be explained by 
%{\color{green}the difference in the outflow properties}
differences in the outflow properties
of individual sources.
%
%We also demonstrate that 
%can be reproduced in the case of 
% that these photon acceleration mechanism can
%account the observed Band spectrum.

\end{abstract}

%\keywords{gamma ray bursts}

\section{INTRODUCTION}

Gamma-ray Bursts (GRBs) are the most powerful explosions in the Universe.
Prompt emission of GRBs are mainly observed in the energy range
of 10 keV - MeV and   show rapid time variability in their lightcurves.
Their spectra are often modeled by an empirical  'Band' function
\citep{BMF93,PBM00,KPB06,KGP08}, which is a
 smoothly jointed
broken power-law, whose physical origin is not yet identified.
%Althogh number of models have been proposed,
%the emission mechanism still remains unclear. 
The typical low and high energy photon indices are distributed 
around $\alpha_{\rm ph} \sim -1$ and $\beta_{\rm ph} \sim - 2.5$, respectively, while
the spectral peak (break) energy is distributed around $E_{\rm p} \sim$ a few $100~{\rm keV}$.

The most widely discussed model for the prompt emission mechanism
 is the internal shock model
\citep{RM94,SP97}.
In this model,
it is assumed that
relativistically moving shells emanate from the central engine
with diverse velocities, and shocks form  due to 
collisions of the shells.
Shocks are accompanied by particle acceleration,
and eventually
gamma-rays are produced by relativistic electrons
via synchrotron radiation.
The internal shock model can naturally 
explain the observed rapid time variability in the lightcurve
and non-thermal nature of the spectra.
However, it is known that the model suffers from poor
radiation efficiency,
since only the kinetic energy associated with the relative motion of
 shells can be released \citep{KPS97, LGC99, GSW01, KMY04}.
This contradicts with observations which show high efficiencies of
a few tens of percent \citep{FP06, ZLP07}.
Another difficulty is  the low energy spectral index ($\alpha_{\rm ph}$).
A non-negligible fraction of observed GRBs show hard spectra
at low energies which cannot be explained by 
 synchrotron emission \citep{CLS97, PBM98, GCG03}.

These well-known difficulties in the internal shock model
have lead researchers to reconsider 
photospheric emission
\citep[e.g.,][]{T94, EL00,
MR00, RM05, LMB09, B11, PR11, MNA11, NIK11, RSV11, XNH12, BSV13, LPR13, LMM13},
which is a natural consequence of
 the original fireball model \citep{G86,P86}.
In this model, prompt gamma-rays 
are released at the photosphere 
when the fireball becomes optically thin.
%In this model, prompt gamma-ray photons are emitted when the 
% initially optically thick outflow becomes transparent as they expand.
The peak energy of the spectra is determined by the 
temperature at the photosphere. %where the photons decouple from the matter.
This model has an advantage in that
high emission efficiencies  can be naturally achieved.
Strong observational support for this scenario
has been provided by the recent detection of quasi-thermal emission
by the Fermi Large Area Telescope (LAT), notably GRB 090902B
\citep{AAA09, RAZ10, RPN11, PZR12}.

However, while there are many advantages, 
 the photospheric emission model must overcome several difficulties.
The major flaw is 
 reproduction of the observed  broad non-thermal spectra,
since the photons that originate from
regions of very high optical depths are inevitably thermalized.
Photospheric emission can, in principle, explain 
the hard low-energy spectra that cannot be accounted for by  synchrotron emission,
by considering a superposition of
emission components which have different temperatures
(multi-color Blackbody).
%\citep[multi-color Blackbody; e.g.,][]{LPR13}.
This is because the Rayleigh-Jeans part of the emission component 
has a much harder photon index ($\alpha_{\rm ph} = 1$) than
those inferred in all observed GRBs.
It is noted, however, that 
the existence of outflow models that can naturally
 account for 
the low energy spectra %well below the peak energy $E_{\rm p}$
is still debated.
Recent theoretical studies which explore peak energies of 
photospheric emissions 
based on spherical (one-dimensional) outflow  models
 \citep[e.g.,][]{L12, B13, VLP13}
suggest that it is  difficult to
regulate the temperature of the emissions to
be well below the typical observed peak energies $E_{\rm p}$,
particularly when dissipation is present 
\citep[but see, e.g.,][for a mechanism that may
overcome this difficulty]{LG13}. 
Although dissipative processes are not considered,
an outflow structure in two-dimensions  suggested by \citep{LPR13}
may be a key ingredient to resolve this difficulty.
%a possible scenario that can overcome the difficulty has 
%been suggested by \citet{LPR13}.
%{\color{green}In their study, %They modeled
%an outflow %
%that have  Lorentz factor gradient in
%the lateral direction %at the outer region 
%was modeled based on the hydrodynamical simulation of
%the jet propagation %within a progenitor star 
%\citep{ZWM03}.}
Based on hydrodynamical simulations of
jet propagation \citep{ZWM03}, \citet{LPR13} modeled an
outflow with a Lorentz factor gradient in the lateral direction.
By solving the transfer of photons within the jet, they indeed 
 showed that the low energy spectra can be
reproduced by the multi-color temperature effects
due to the superposition of the photons released 
from different lateral positions that have various Lorentz factors.
 %having large Lorentz factor gradients.
%\citep[]{L12, B13, VLP13}
%One way to overcome the difficulty to 
%
%,LG13, [see] aa}[]{LG13}.
%the origin of the low temprature components that can
%naturally account for the low energy spectra is still debated,
%
% emission components having peaks at low energies
%(temperatures) are difficult to produce, 
%in particular when dissipations are present below
%the, photosphere \citep[][]{L12, B13, VLP13,LG13}.
%}

On the other hand, 
the high energy
part of the observed
spectra %index ($\beta_{\rm ph}$) 
is difficult to reproduce by the super-position of 
thermal emissions, since an unreasonably high temperature
must be realized within the outflow.
To overcome this difficulty, 
dissipative processes
around the photosphere have been suggested in previous studies.
Various mechanisms
such as
shocks \citep{PMR05, PMR06, IMT07, LB10},
 magnetic reconnection \citep{G06, GS07, G08} 
and proton-neutron nuclear collisions \citep{B10, VBP11} have been proposed.
All models are accompanied by
 the generation of relativistic electrons (and positrons) 
\footnote{
It  should be noted that the Fermi acceleration 
of charged particles  at 
 shocks below the photosphere, which is occasionally 
adopted in previous studies,
 is highly unlikely to be the origin of the relativistic electrons,
since the typical width of the shock transition,
 a few Thomson mean free paths, is much larger than any
kinetic scale involved by many orders of magnitudes \citep{LB08, KBW10, BKS10}.
}
 that upscatter the thermal photons to
 create the non-thermal spectra  \citep{G12}.
%by Compton scatterings \citep{G12}.
%It is noted, however, the 
%Hence, %in this scenario
%efficient energy dissipation 
%copious amount of 
%energy must be transferred to the non-thermal particles.
However,
 it is quite uncertain whether
%{\color{green}the dissipative processes can operate efficiently to
%supply copious amount of energy
%to the relativistic electrons (and positrons)
% which is comparable to that of the dominant thermal energy 
%so that the reproduction of  the observations is possible.}
dissipative processes can operate efficiently enough to
deposit in relativistic electrons (and positrons) the copious amounts
of energy needed to reproduce observations.

Although the details are not well studied,
Fermi-like acceleration of photons at the shocks formed below the photosphere
due to bulk Compton scattering 
can be regarded as
an alternative mechanism for producing high energy non-thermal spectra 
\citep{E94}.
Presence of relativistic electrons (and positrons) 
is not required in this scenario, 
%Strong dissipative processes are not required in this scenario,
since the energy of the  background fluid can be directly
transferred to the photons as they propagate.
% photons gain energy directly from the background fluid
%as they propagate. 
Shocks formed below the photosphere
are inevitably mediated by radiation \citep{LB08}.
The properties of relativistic radiation mediated shocks have been 
explored in various astrophysical contexts
(e.g., shock breakouts) including GRBs.
The most detailed and fully self-consistent analysis 
of relativistic radiation mediated shocks have been performed by
\citet{BKS10} \citep[see also][]{KBW10}.
Although not as sophisticated as the above study,
\citet{BML11} explored the properties of  photon acceleration 
in the context of shocks within  GRB jets.
These studies have shown that the existence
of accelerated photons  above the thermal peak energy
is an inherent feature in these shocks.
%
%Although not as sophisticated as the above study,
%\citet{BML11} explored the properties of the photon acceleration 
%in the context of shocks within  GRB jets. 
%emissions accompanied by subphotospheric shocks in
%By performing %two-dimensional
%%%Monte-Carlo test particle
%simulations, they demonstrated that, the hard non-thermal tail is likely to 
%be present in the resultant spectra for the conditions realized
%if the shocks are present in the subphotospheric regions.
%\bl{
%However, 
%detail comparisons with observations were not considered and
%the obtained spectra are far from the typically observed one.
However, all the above studies assume 
a one-dimensional stationary planar shock in their analysis, 
and the obtained spectra are far from those typically observed.
Therefore,
for a more accurate estimation,
a global photon transfer calculation incorporating the multi-dimensional
structure of the jet must be performed.
\citet{IOK11} explored the properties of shocks
that are formed in radiation dominated jets. 
They claimed that photons 
which are accelerated at the shock
can generate a spectra which are close to the  'Band' function.
However, since their argument is based on a
simple analytical framework in which the transfer of photons is
not solved, it is obvious that
detailed calculations are required to form firmer conclusions.

While the previous studies focused on the  shocks, 
any form of large velocity gradients that are present in the 
outflow can give rise to photon acceleration.
For instance, a
Fermi-like mechanism can also operate
when velocity shear exists in the transverse direction of the outflow,  
just as in the case of the acceleration of the cosmic rays
\citep[e.g.,][]{JKM89, O90, O98}.
Indeed, 
GRB jets are likely to posses
rich internal velocity structures
in the transverse directions.
For example, numerical studies
which explore the launching mechanism of the relativistic jets
suggest that the produced jets have transversely  stratified structures
in which the bulk Lorentz factor increases towards the jet axis
\citep{Mc06, NTM07, N09, MB09, N11}.
Since the highly magnetized jets seen in these simulations are cold,
 subsequent magnetic field dissipation must 
occur  in order to  produce a hot fireball which leads to an
efficient photospheric emission.
Although the flow structure after the dissipative process is highly 
uncertain, the stratified structure that
originated at the central engine may remain even at this stage.
Moreover,
rich velocity structure can also develop
after the launching phase as the jets drill through the progenitor star
envelopes.
A large number of 
hydrodynamic simulations regarding the jet propagation
show  that large velocity gradients
 appear within the jet (in radial as well as transverse directions)
due to the interaction with the stellar envelope
\citep{ZWM03, MYN06, MLB07, LMB09, MNA11, NIK11}.
Notably,
 large velocity gradient in the transverse directions produced by
 the strong recollimation shocks seen in these studies 
will likely  provide an
 efficient acceleration site for photons as pointed out by
\citet{L12}.

Motivated by this background,
we explore the photon acceleration within 
a jet with velocity structures in the transverse direction %(stratified jet)
and its effect on the resulting spectra of the photospheric emissions
by solving the transfer of photons.
As mentioned earlier, mainly focusing on the origin of the 
low energy spectra, \citet{LPR13} carried out a similar study.
By performing a photon transfer calculation,
they evaluated
the photospheric emission from a relativistic jet
that has a  continuously  decaying velocity profile 
in lateral direction $\Gamma \propto \theta^{-p}$
at the outer region.
However, while they succeeded in reproducing the 
spectra below the peak energy,
the imposed
velocity gradient was not large enough to trigger
efficient photon acceleration
and, therefore, 
the high energy part
was  too soft compared to the Band spectra.
%(although the sign of photon acceleration was seen in some cases)..
%
%}
With the aim to investigate the cases with 
efficient photon accelerations, 
we focus on jets that have a
velocity shear in 
the transverse direction.
This can be considered as a limiting case
of $p=\infty$ in \citet{LPR13} work, a parameter region which was
not explored.
%Contrary to their study,
%in the aim
%to investigate the cases with efficient photon accelerations,
%we focus on jets that have a velocity shear in 
%the transverse direction.
As a first step, this study considers a
simple stratified two-component jet structure, wherein a
highly relativistic spine outflow is surrounded by a wider and
less relativistic sheath outflow.
We show, in particular, that the
high-energy part of the Band spectra (photon index $\beta_{\rm ph}$) can indeed
be reproduced by  photons accelerated at the shear layer.
In addition, 
we also demonstrate that the low-energy 
part of the spectra (photon index $\alpha_{\rm ph}$)
 can be reproduced consistently
within the framework of the present study
when time evolution of the outflow is considered.
%
%\bl{
%In reality, the velocity shear is not stable
%due to the effects such as Kelvin-Helmholz instability,
%and velocity should have continous velocity profiles.
%This
%}

The paper is organized as follows. 
In \S\ref{model}, we describe our model and numerical procedures.
We present our main results in \S\ref{result}.
Discussions of the implications of our results are given in
 \S\ref{discussions}.
The summary of our main findings are given in \S\ref{summary}.

%Here, we do not consider shocks but assume 
%spine-sheath jet and study the 
%case of photon acceleration when
%shear flow is present in the
%lateral direction of the outflow. % (spine-sheath structure)

%it is shown that Band spectra can successfully
%reproduced in this scenario.
% 

\section{MODEL AND METHODS}
\label{model}

\begin{figure}[ht]
\begin{center} 
\includegraphics[width=7cm]{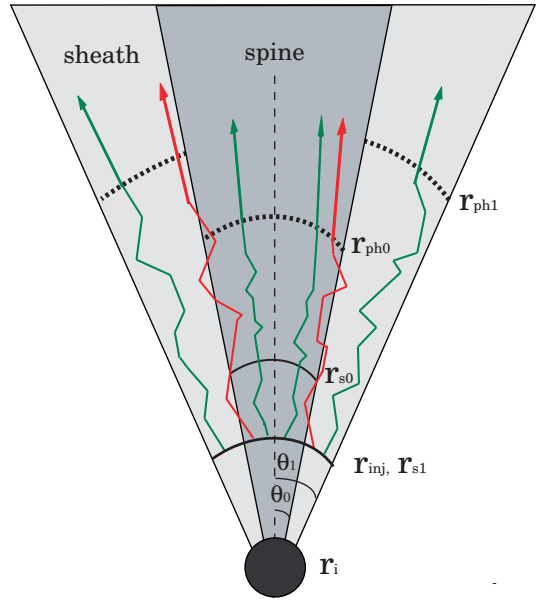}
\caption 
{Schematic picture of the employed model.
 A fast spine jet is embedded in a slower sheath outflow. Spine and
 sheath start to accelerate at a radius $r_{\rm i}$, and
 the acceleration continues up to $r_{\rm s0}$ and $r_{\rm s1}$
 in the spine and sheath region, respectively.
 Since the dimensionless entropy of the spine $\eta_0$ 
 is larger than that of the sheath $\eta_1$,  
 the saturation radius and the terminal Lorentz factor 
 of the spine ($r_{\rm s0} = \eta_0 r_{\rm i}$ and $\Gamma_0 = \eta_0$)
 are larger than that of the sheath 
 ($r_{\rm s1} = \eta_1 r_{\rm i}$ and $\Gamma_1 = \eta_1$).
 The photospheric radius of the spine $r_{\rm ph0}$ is smaller than that of the sheath $r_{\rm s1}$.
 In our model, thermal 
photons are injected at a the saturation radius of the sheath $r_{\rm inj} = r_{\rm s1}$, and the transfer is solved up to the radius in which
   the optical depth is much lower than unity.
 Photons which
 cross the boundary layer of the spine and sheath
 multiple times  gain energy and produce a non-thermal tail in the resultant 
 spectrum.} 
\label{modelpic}
\end{center}
\end{figure}

In the present study,
we evaluate the photospheric emissions from
an ultra-relativistic outflow which has a
spine-sheath jet structure.
%by solving the transfer of photons within the jet.
%
The spine is defined as a region of conical outflow with a
half-opening angle $\theta_0$, while the sheath is a region
of outflow which surrounds the spine
and extends up to an angle of $\theta_1(>\theta_0)$ (see Fig. \ref{modelpic}).
Each region has different fluid properties and a steady radial outflow
is assumed.

%
%In GRB jets, opacity of photons is strongly dominated by the scatterings 
%with electrons.
%Therefore,
%we neglect the absorption process and
%only consider the scattering process in our calculations.
%
%The assumed background fluid properties
%is similar to that considered in \citet{PR11} and \citet{B11}.
%The major difference is that, while spherical symmetry is 
%assumed in these studies, we assume axisymmetry and
%consider structure in
%the lateral ($\theta$) direction of the outflow.

%\subsection{Photon Transfer in Spine-Sheath Jet}
\subsection{Fluid Properties of Spine-Sheath Jet}
\label{PTSJ}

\subsubsection{Radial evolution}
\label{radial}

%Here we describe how the radial evolution of 
%fluid propeties are determined

%The spine is defined as a region of conical outflow with
%half-opening angle of $\theta_0$, while the sheath is a region
%of outflow which surrounds the spine
%and extends up to an angle of $\theta_1(>\theta_0)$ (see Fig. \ref{modelpic}).
%In each region, a steady radial outflow is assumed.
Under the assumption
that the energy density in the magnetic field and the 
dissipated kinetic energy are sub-dominant,
%that the effects of magnetic field and energy dissipations are passive,
the fluid properties of the spine and sheath regions are 
described by the  standard adiabatic ``fireball'' model
\citep[e.g.,][]{P04, M06}.
Here we give a brief review of the fireball model and 
show how the radial evolution of the
Lorentz factor (velocity), $\Gamma$,  and
electron number density, $n_e$, which are
necessary background fluid
 information for evaluating the photon transfer, is determined.

%We assume a steady radial outflow described by
%the standard ``fireball'' model
%\citep[see, e.g.,][]{P04, M06} for both regions 
%in determining the electron number density and the bulk Lorentz factor.
%
The dynamics of the fireball can be characterized by three
independent parameters, which are the initial fireball radius, $r_{\rm i}$,
the kinetic luminosity, $L$,  and the
 dimensionless entropy 
 which characterizes the baryon loading,
  $\eta \equiv L/\dot{M}c^2$, where $\dot{M}$ and $c$ are
 the mass outflow rate and the speed of light, respectively.
Initially, the fireball accelerates as it expands 
by converting its internal energy into kinetic energy
(acceleration phase). 
As a result,
the bulk Lorentz factor of the flow increases with radius, $r$,  as  
$\Gamma(r) \simeq (r/r_{\rm i})$.
%
%There is a critical value in 
%
%Assuming that pair production is negligible,
If $\eta$ is larger than the critical value
given by $\eta_c = (\sigma_{\rm T} L / 8\pi r_{\rm i} m_p c^3)^{1/4}$
\citep[][]{P04, M06},
the fireball becomes transparent to radiation during the acceleration phase,
where $\sigma_T$ and $m_p$ are the
 Thomson cross section and proton rest mass, respectively.
On the other hand, if $\eta < \eta_c$, 
the fireball becomes transparent
after the acceleration has ceased and 
the flow asymptotically approaches a constant Lorentz factor $\Gamma = \eta$ 
(coasting phase).
In this case, 
the acceleration continues up to the 
 saturation radius,
 $r_{\rm s} = \eta r_{\rm i}$.
%and, thereafter, the flow has a constant Lorentz facto 
%and the saturation radius $r_{\rm s} = \eta r_{\rm i}$.
In the present study, we focus on the latter case and
 consider the  parameter space
in which $\eta < \eta_c$ is satisfied
in both the spine and sheath regions. 
%Therefore, the matter becomes transparent to radiation
%when both regions are in coasting phase.
Hence, the radial evolution of Lorentz factor is given by
\begin{eqnarray}
\label{Gamma}
\Gamma(r) =   \left\{ \begin{array}{ll}
      \frac{r}{r_{\rm s}} &~~
         {\rm for}~~ r \leq r_{\rm s}  ,  \\
     \eta  &~~
         {\rm for}~~  r > r_{\rm s}  . \\
             \end{array} \right. 
\end{eqnarray}
%focus on
%the range of radii in which both spine and sheath are in 
%the coasting phase ($r \geq r_{\rm s}$).
%
%Hence, spine and sheath expand in the adial direction
%and have constant Lorentz facotor with different values
%in the calculation range.
%
%Hereafter, we use the terminal Lorentz factor 
%$\Gamma (=\eta)$ as a parameter of fireball model 
%instead of $\eta$, merely for convenience.
%

In the relativistically expanding outflow,
the electron number density in the comoving frame
is given by
%$n_e(r) = \dot{M}/4 \pi r^2 m_p \Gamma \beta c
%= L / 4 \pi r^2 m_p \eta \Gamma \beta c^3$, 
\begin{eqnarray}
\label{ne}
n_e(r) = \frac{\dot{M}}{4 \pi r^2 m_p \Gamma \beta c}
 = \frac{L}{4 \pi r^2 m_p \eta \Gamma \beta c^3},
\end{eqnarray}
%= L / 4 \pi r^2 m_p \Gamma^2 \beta c^3 \propto r^{-2}$,
where $\beta$ is the velocity of the flow normalized by 
the speed of light. 
Here, we assumed that there are no strong dissipative processes
in the outflow which may create copious pair plasma.
Since we consider the case of $\eta < \eta_c$, the
electron number density decreases with radius as
$n_e \propto r^{-3}$ below the saturation radius ($r\leq r_{\rm s}$) and
as $n_e \propto r^{-2}$ at larger radii ($r> r_{\rm s}$).

Given the electron number density and bulk Lorentz factor
of the flow,
the optical depth to Thomson scatterings for
the photons propagating in the {\it radial} direction
to reach infinity can be evaluated as
%$\tau(r) = \int^{\infty}_r \sigma_T n_e(r') \Gamma (1-\beta) dr' 
%\simeq \sigma_T n_e(r) r/ 2\Gamma $,
%\begin{eqnarray}
%\tau(r) = \int^{\infty}_r \sigma_T n_e(r') \Gamma(1-\beta) dr' 
%\simeq \frac{\sigma_T n_e(r) r}{2\Gamma} = \frac{r_{\rm ph}}{r},~~
%r_{\rm ph} = \frac{\sigma_{\rm T} L}{8 \pi \eta^3 m_p c^3},
%\end{eqnarray}
\begin{eqnarray}
\label{tau}
  \tau(r) &=&  \int^{\infty}_r \sigma_{\rm T} n_e(r') \Gamma(r')(1-\beta(r')) dr' \nonumber \\
&\simeq&
  \left\{ \begin{array}{ll}
    \frac{r_{\rm ph}}{r} \left[1+\frac{1}{3}
         \left\{ 
       \left(\frac{r_{\rm s}}{r} \right)^2
     -  \left(\frac{r}{r_{\rm s}} \right)^2 \right\} \right]  &~~
         {\rm for}~~  r \leq r_{\rm s}  , \\
      \frac{r_{\rm ph}}{r} &~~
         {\rm for}~~ r > r_{\rm s}  ,

  \\
             \end{array} \right. 
\end{eqnarray}
\begin{eqnarray}
\label{rph}
r_{\rm ph} = \frac{\sigma_{\rm T} L}{8 \pi \eta^3 m_p c^3},
\end{eqnarray}
where we have assumed $\Gamma \gg 1$.
Here, $r_{\rm ph}$ is the photospheric radius
which corresponds to the radius
where the optical depth becomes unity ($\tau=1$).
%
%Note that the above evaluation of the optical depth can only 
%be applied for the photons propagating in radial direction.
%In an ultra-relativistic outflow, the optical depth 
%is quite sensitive the direction of the photon.
%%The optical depth of the photon propagating in relativistic flow
%%is quite sensitive to the direction of the photon propation.
%In calculating the photon propagation, we properly take into
%account of this effect as described in \S\ref{MCcode}.

%Note that the saturation radius as well as photospheric
%radius is different for the spine and sheath since
%we inpose different values in the parameters in the fireball
%for each region.

%The radius where the 
%(photospheric radius) is derived from the above equation 
%as a radius which satisfies $\tau = 1$ and is given by
%$R_{\rm ph} = \sigma_{\rm T} L / 8 \pi \Gamma^3 m_p \beta c^3 
% \approx  9.2 \times 10^{11}L_{53}\Gamma_{400}^{-3}~{\rm cm}$
%\begin{eqnarray}
%R_{\rm ph} = \frac{\sigma_{\rm T} L}{8 \pi \Gamma^3 m_p \beta c^3} 
% \approx  9.2 \times 10^{11}L_{53}\Gamma_{400}^{-3}~{\rm cm},
%\end{eqnarray}
% where $L_{53}=L/10^{53}~{\rm erg/s}$ and $\Gamma / 400$.
%It is noted that 
%using the photospheric radius, the optical depth can be
%expressed as $\tau(r)=R_{ph}/r$.

\subsubsection{Spine-sheath structure}
\label{SPst}

%While the fluid quantities ($n_e(r)$ and $\Gamma(r)$)
%are uniform within each regions,
The radial profiles of the fluid quantities ($n_e(r)$ and $\Gamma(r)$)
in the spine  and sheath regions are different, since 
we impose different values on their
fireball parameters
($r_{\rm i}$, $L$ and $\eta$). % in the fireball for each region.
In the present study,
while we assume the same value for $r_{\rm i}$ in both 
regions, the dimensionless entropy of the spine, $\eta_0$,
is taken to be larger than that of the sheath, $\eta_1$.
Hence, the saturation radius in the spine region,
$r_{\rm s0}= \eta_0 r_{\rm i}$, is larger than
that of the sheath, $r_{\rm s1}= \eta_1 r_{\rm i}$, and
the terminal Lorentz factor of the former
($\eta_0$) is larger 
than that of the latter ($\eta_1$).
Regarding the evolution of the Lorentz factor
in the spine and sheath,
they have equal values 
($\Gamma_0(r) = \Gamma_1(r) = r/r_{\rm i}$) up to $r=r_{\rm s1}$.
At larger radii ($r>r_{\rm s1}$), 
velocity shear begins to develop and the
difference in the Lorentz factor
increases with radius up to $r_{\rm s0}$, since
the spine is in the acceleration phase while the
sheath is in the coasting phase ($\Gamma_1 = \eta_1$). 
Thereafter, the spine also enters the coasting phase ($\Gamma_0 = \eta_0$), and
the difference in the Lorentz factor is constant.
In determining the kinetic luminosities of the spine, $L_0$, and
sheath, $L_1$, we assume that the mass outflow rate
is equal in both regions ($L_0 / \eta_0 = L_1 / \eta_1$).
Therefore, kinetic luminosity
of the spine is larger by a factor of $\eta_0/\eta_1$.
Under these assumptions, 
 the photospheric radius 
in the sheath, $r_{\rm ph1}$, is larger than that in the spine,
 $r_{\rm ph0}$, by a factor of $(\eta_0/\eta_1)^2$ (see equation (\ref{rph})).
A schematic picture of the employed model is given in Fig. \ref{modelpic}.

Hereafter,
the quantities corresponding to the spine and sheath
regions are denoted by subscript $0$ and $1$, respectively.

%For given, $\Gamma_0$ and $\Gamma_1$, the
%kinetic luminosities in the spine region, $L_0$, and
%sheath region, $L_1$, are determined
%in the way that the mass outflow rate become 
%equal in both regions ($L_0 / \Gamma_0 = L_1 / \Gamma_1$).

\subsection{Photon Transfer in a Spine-Sheath Jet}
\label{Phtr}
%\subsubsection{Treatment of Photon transfer}    

Having determined the background fluid properties ($\Gamma$ and $n_e$),
we evaluate the resultant photospheric emission
by solving the propagation of photons which are injected 
far below the photosphere.
%within the flow.
%
The photon transfer is evaluated by performing a three-dimensional
test particle Monte-Carlo simulation.
In GRB jets, opacity of photons is strongly dominated by the scatterings 
with electrons.
Therefore,
we neglect the absorption process and
only consider the scattering process by the electrons in our calculations.
Furthermore
 we do not take into account
the thermal motion of the electrons in evaluating the scattering
 for simplicity.

%and assume that the electrons have zero temperature.

\subsubsection{Initial condition}  
\label{Iniph}  

Initially, the photons are injected within the jet 
at the surface of a fixed radius where the velocity
 shear begins to develop $r_{\rm inj} = r_{\rm s1}$.
For the cases considered in this study,
$r_{\rm inj}$ is always located far below the 
photosphere ($\tau(r_{\rm inj})\gg 1$). Therefore,
a tight coupling between the photons and matter is expected.
%
%Since the optical depth is sufficiently larger than unity
%at the injection radius ($\tau(r_{\rm inj})\gg 1$)
%for the cases considered in the present study,  
%the photons are strongly coupled with the matter and
%expected to be fully thermalized.
%
For this reason, we can safely
assume that the photons
have an isotropic distribution
with energy distribution  given by a Planck distribution in the comoving frame.
%
%As in the case of the backgound fluid,
%the comoving temperature of the injected photons is
%determined based on the  fireball model.
%As for the comoving temperature of the photons,
%we adopt the corresponding value determined by
%the fireball model at $r = r_{\rm inj}$ which is given as
% $T_{\rm inj} = (L/4\pi r_{\rm i}^2 c a )^{1/4} ( r_{\rm inj}/r_{\rm i})^{-1}$,
%where  $a$ is the radiation constant.
%its radial evolution is given by
%is determined by the corresponding temperature
%of the  fireball model
%in which the radial evolution of the temperature is given by
%When the coupling between the photon and the matter is strong,
%the photons are subject to adiabatic loss due to the 
%expansion of the flow,
%and the comoving temperature decreases with radius as
According to the fireball model, the radial evolution of the 
comoving temperature is given by
\begin{eqnarray}
\label{Tev}
T'(r) =   \left\{ \begin{array}{ll}
      \left( \frac{L}{4\pi r_{\rm i}^2 c a} \right)^{1/4}
      \left( \frac{r}{r_{\rm i}} \right)^{-1}
 &~~
         {\rm for}~~ r\leq r_{\rm s}  ,  \\
 \left( \frac{L}{4\pi r_{\rm i}^2 c a} \right)^{1/4}
      \left( \frac{r_{\rm s}}{r_{\rm i}} \right)^{-1}
      \left( \frac{r}{r_{\rm s}} \right)^{-2/3}
       &~~
         {\rm for}~~  r> r_{\rm s}  , \\
             \end{array} \right. 
\end{eqnarray}
where  $a$ is the radiation constant.
Hence,
we adopt the temperature at the corresponding radius
given by above equation  $T'_{\rm inj} = T'(r_{\rm inj})$
for the comoving temperature of the injected photons.
% the photons are injected at the sphere of fixed radius $r_{\rm inj}$
%with radiation intensity
%given by the blackbody having temperature 
%of $T_{\rm inj} = T(r_{\rm inj})$.
%
%
While the photons are isotropic in the comoving frame,
they are strongly beamed in the laboratory frame
 due to the Doppler boosting effect.
%In the laboratory frame, the injected photons are
%strongly beamed due the doppler boosting and the corresponding  
%the radiation intensity given by 
% the photons are injected at the sphere of fixed radius $r_{\rm inj}$
%with a radiation intensity given by
Due to this effect,
the radiation intensity  of the blackbody emission
 in the laboratory frame
is given by
\begin{eqnarray}
\label{Iin}
I_{\nu,{\rm inj}}(\nu) = {\cal D}(\Gamma_{\rm inj}, \theta_v)^3
 B_{\nu}(T'_{\rm inj}, \nu/{\cal D}(\Gamma_{\rm inj}, \theta_v)),
\end{eqnarray}
where
% $T_{\rm inj}=T(r_{\rm inj})$
 $\Gamma_{\rm inj} = \Gamma(r_{\rm inj})$ 
%are the comoving temperature
is the bulk Lorentz factor of the flow
 at $r=r_{\rm inj}$ determined from equation (\ref{Gamma}).
%Hence, for the temperature of the injected photons,
%we employ $T_{\rm inj}=T(r_{\rm inj})$.
%
%The temperature of Plank distribution is determined
%by the corresponding
%comoving temperature of the fireball model at the injection radius $r_{\rm s1}$ which
%is given as 
%$T_{\rm inj}=(L_0/4\pi r_{\rm i}^2 c a)^{1/4}(r_{\rm s1} / r_{\rm i})^{-1}$ and $T_{\rm inj}=(L_1/4\pi r_{\rm i}^2 c a)^{1/4}(r_{\rm s1} / r_{\rm i})^{-1}$ for the spine and sheath, respectively, where
%$a$ is the radiation constat of the sheath by a factor of $(\eta_0/\eta_1)^{1/2}$, since
%the kinetic luminosity is higher by a factor of $(\eta_0/\eta_1)^2$
%
%Based on the assumptions, 
%the photons are injected the sphere of $r=r_{\rm inj}$ with
%a radiation intensity given by
%$I_{\nu}(\nu) = {\cal D}(\theta_v)^3
% B_{\nu}(T_{\rm inj}, \nu/{\cal D}(\theta_v))$
%in the laboratory frame.
Here, $B_{\nu}(T',\nu) =
2 h \nu^3 c^{-2} [{\rm exp}(h\nu)/k_{\rm B} T' - 1]^{-1}$ is the Planck function,
where
$h$ and
$k_{\rm B}$
are the Planck constant and the Boltzmann constant, respectively,
and
$D(\Gamma, \theta_v)=[\Gamma(1-\beta {\rm cos}\theta_v)]^{-1}$ is the 
Doppler factor, where
$\theta_v$ is
 the angle between
the photon propagation direction and the fluid velocity direction
(radial direction).
In our calculations, 
the initial propagation direction and frequency of
the injected photons are drawn
from a source of photons given by the above equation.

It is emphasized that the results of our calculation are insensitive to
the assumed position of the injection radius as long as
$r_{\rm inj} \leq r_{\rm s1}$ is satisfied.
This is because, at a radius far below the
photosphere ($\tau(r) \gg 1$),
the photon energy distribution
evaluated by solving the photon transfer 
does not
deviate from the Planck distribution if velocity shear is not present (see next section),
and its temperature evolution is well described by 
equation (\ref{Tev}).  
The temperature of the injected photons in the spine, $T'_{\rm inj0}$, 
is higher than that in the
sheath, $T'_{\rm inj1}$, by a factor of $(\eta_0/\eta_1)^{1/2}$, since
the kinetic luminosity of the former
 is higher by a factor of $(\eta_0/\eta_1)^2$.
Correspondingly,
the luminosity of the injected photons in the laboratory frame
($L_{\rm inj} \propto r_{\rm inj}^2 \Gamma_{\rm inj}^2 T_{\rm inj}^{'4}$)
in the spine is higher than that of the sheath
by a factor $\sim (T'_{\rm inj0}/{T'}^{4}_{\rm inj1}) \sim (\eta_0/\eta_1)^{2}$. 
%
%The injection rate of the photon 
%is determined %also determined from the fireball model
%so that the luminosity of photons in the lab frame 
%becomes identical to that of the fireball model at the injection radius
%given by $L_{\gamma}(r_{\rm inj})=L(r_{\rm inj}/r_{\rm s})^{-2/3}$.
%It is noted that the temperature and the injection rate
%at the injection radius
%is different between the spine and sheath since the assumed parameters of
%the fireball is different.
%

%The above initial condition for the photons are justified for
%the following two reasons.
%One is that, since the injection radius employed in the 
%the present study is deep below the photosphere 
%($\tau(r_{\rm inj}) \gg 1$), the photon are expected to
%be fully thermalized. % if there are no shear velosity (see next section).
%Another is that, since  velosity shear is not present 
%below the injection radius,
%photons below the injection radius
%are not subject to the photon acceleration which 
%may lead to severe deviation from thermal distribution (see next section).

%Another is that, although shear velosity is present
%below the injection radius
%in the range of  $r_{\rm s1} \leq r \leq r_{\rm inj}(=r_{\rm s0})$,
%the deviation from the thermal distribution due
%to the velosity difference can be neglected,
%since the length and strength the shear flow
%in this radius are not
%sufficient to affect the resultant photon spectra
%signficantly (for detail, see next section). 

\subsubsection{Boundary conditions}

After the photons are injected, we track their path within
the jet in  three-dimensions
by  performing Monte-Carlo simulations (\S\ref{MCcode}) 
until they reach the outer or inner boundary
of the calculation.
%The transfer of the injected photons is solved by performing 
%Monte-Carlo simulations (see \S\ref{MCcode} for detail)
% until
% they reach the outer or inner boundaries
%of the calculation.
Using spherical coordinates ($r$,$\theta$,$\phi$), %In radial direction 
the outer boundary $r$ 
is set at a radius $r_{\rm out}=500r_{\rm ph0}$
where the photons can be safely
considered to have  escaped since the optical depth is 
$\tau(r_{\rm out})=2\times10^{-3} \ll 1$.
While there is no boundary in the $\phi$ direction,
outer boundary in the $\theta$ direction is set 
at $\theta_{\rm out}=\theta_1$ which corresponds
to the edge of the whole jet.
As for the inner boundary, 
we adopt a radius slightly below the injection radius
 $r_{\rm in}=0.5r_{\rm inj}$.
%and at angle $\theta_{\rm out}=\theta_1$ which corresponds
%to edge of the whole jet, while the 
%inner boundary is fixed below 
%the  injection radius $r_{\rm in}=0.5r_{\rm inj}$.
For photons which have reached the outer boundaries,
we assume that they escape freely to $r=\infty$ without being
scattered or absorbed.
On the other hand, we assume that the photons are simply absorbed
in the inner boundary.
It is noted, however, that
the fraction of absorbed photons is negligible,
since  most of the photons in ultra-relativistic outflows 
are strongly collimated due to the relativistic beaming effect
 and essentially streamed outward \citep[e.g.,][]{PR11,B11}.

%In the present study,
The spectra of the emission are evaluated from the 
photons which have reached the outer boundaries.
Due to the relativistic beaming effect,
these photons
 are highly anisotropic and
 mostly concentrated within a
cone of half-opening angle $\sim \Gamma^{-1}$ in
the direction of the fluid velocity (radial direction) at
the last scattering position. 
Hence, the observed emission spectra
depend significantly on the angle between the
direction to the observer and the jet axis, $\theta_{\rm obs}$. 
In the present study,
we evaluate the spectrum for observers
located at direction $\theta_{\rm obs}$
by recording all photons which have reached the outer 
boundary that are propagating in direction 
within a cone of half-opening angle $\Gamma^{-1}$,
which is small enough to regard that
the emission is uniform within the cone.
From the recorded photon flux, 
we calculate the isotropic equivalent luminosity 
by multiplying the photon flux by a factor  $4\pi/d\Omega$,
where
 $d\Omega=2\pi[1-{\rm cos}(1/\Gamma)]$ is
% $d\Omega=2\pi \int^{1/2\Gamma}_0 {\rm sin}\theta d\theta$ is
the solid angle of the cone.

%Since the most of the photons are emitted within 
%a cone with half-opening angle $\sim \Gamma^{-1}$
%in the direction
%of the fluid velosity (radial direction) due to
%the relativisitic beaming effect, the emissions are highly anisotropic
%and varies largely depending on the angle between
%direction to the observer and jet axis, $\theta_{\rm obs}$. 
%In evaluating the spectrum for observer
%located at direction of $\theta_{\rm obs}$,
%we record all pariticles which have reached the outer 
%boundary that are propagating in direction 
%within a cone of half-opening angle $(2\Gamma)^{-1}$
%and calculate isotropic equivalent luminosity.

\subsubsection{Monte-Carlo simulation for solving photon transfer}
\label{MCcode}

Here we briefly describe the Monte-Carlo code used
to solve the photon transfer.
As noted previously, we
neglect the thermal motions of the electrons and
 only take into account the scattering processes.
%and neglect their thermal motions. % of the electrons.
%Under the above assumptions, the tranfer of the photons are 
%solved as follows.
Hence, the rest frame of the fluid is equivalent to that of the electrons.
Under the above assumptions, %given the position and 4-momentum of photon,
the propagation of the photons is performed by directly tracking the path
of the individual photons in the three-dimensional space
of the calculation.
Each photon has a specified position  propagation direction and frequency, 
and these quantities are updated by using a uniform random number.

Within the  jet,
the photons travel
along straight paths before they are scattered by the electrons.
Firstly, 
the code determines 
 the distance for the photons to travel
before the scattering %event takes place.
%This is done
by drawing the corresponding optical depth  $\delta \tau$. 
%to the scattering position using random number.
The probability for the selected optical depth 
to be in the range of 
[$\delta \tau$, $\delta \tau + d\tau$] 
is given as ${\rm exp}(- \delta \tau) d\tau$.
%A large number of photon propagation
%is followed by directly tracking the path
%of the individual photons.
%Firstly, %using an uniform random number,
%the code chooses an optical depth for a photon to travel 
%before the scattering takes place.
%
%The probability for the chosen optical depth
%to be in the range of 
%[$\delta \tau$, $\delta \tau + d\tau$]
%is given by ${\rm exp}(- \delta \tau) d\tau$.
%
Then, from the given optical depth $\delta \tau$,
the distance $l$ to the scattering event is
determined from the integration along the straight path
of photons which can be expressed as 
\begin{eqnarray}
\label{taucal}
\delta \tau = \int^l_0 n_e \Gamma(1-\beta {\rm cos}\theta_v) \sigma_{\rm sc} dl, 
\end{eqnarray}
where $\theta_v$ is 
the angle between the direction of fluid velocity and photon.
Here, $\sigma_{\rm sc}$ is the total cross section for the electron 
scattering and is given as 
\begin{eqnarray}
\label{sccross}
\sigma_{\rm sc} =   \left\{ \begin{array}{ll}
      \sigma_{\rm T}
 &~~
         {\rm for}~~ h \nu_{\rm cmf} \leq  100~{\rm keV}  ,  \\ 
%\leq 0.1 m_e c^2  ,  \\
%
%
 \sigma_{\rm KN}
       &~~
         {\rm for}~~  h \nu_{\rm cmf}  >   100~{\rm keV}  ,  \\ 
%0.1 m_e c^2  , \\
             \end{array} \right.
\end{eqnarray}
in our code, where $\sigma_{\rm KN}$ is
the total cross section for  
Compton scattering, and
$\nu_{\rm cmf}$ %and $m_e$ are
is the frequency of the photon in electron (fluid)
comoving frame. % and the electron rest mass, respectively. 
(The frequency $\nu_{\rm cmf}$
 is evaluated by performing a Lorentz transformation
 using local fluid velocity).
Given the distance $l$ from the above equation,
we update the position of the photons to the scattering position
by shifting them from the initial position
with the given distance
 in the initial direction of photon propagation.
Note that, unlike the case of equation (\ref{tau}),
the optical depth calculated by 
equation (\ref{taucal}) is not limited to photons propagating  
in the radial direction.
The path of integration is along the straight path of photons
which can be in an arbitrary direction. 
For a given value of $\delta \tau$,
the distance $l$ strongly depends on the propagation direction of the photons
in the case of a relativistic flow ($\Gamma \gg 1$).
As is obvious from the above equation,
the mean free path of photons
 $l_{\rm mfp} = [n_e \Gamma (1-\beta {\rm cos}\theta_v)]^{-1}$ is
quite sensitive to the photon propagation direction,
since the factor $\Gamma (1-\beta {\rm cos} \theta_v)$ varies
largely from $\sim (2\Gamma)^{-1}$ (for ${\rm cos}\theta_v = -1$) 
up to $\sim 2\Gamma$  (for ${\rm cos}\theta_v = 1$) depending on the 
value of $\theta_v$.
Hence, a  photon tends to travel a larger distance in 
the fluid velocity
(radial) direction since the mean free path of the photon tends to be larger.  
Hereafter, quantities measured in the comoving frame of the
fluid (electron) are denoted by tilde.

%As mentioned in \S\ref{PTSJ}, we consider two component jet
%in which each component is described by fireball model.
%In our code,
In evaluating the integration in equations (\ref{taucal}),
we employ two different methods
 depending on the frequency and
position of the photon.
For photons located above the saturation
radius $r\geq r_{\rm s}$ ($\Gamma = {\rm const}$)
that satisfy $h \nu_{\rm cmf} \leq   100~{\rm keV}$
%0.1 m_e c^2$ 
($\sigma_{\rm sc} = \sigma_{\rm T} = {\rm const}$), % along their path, 
analytical integration can be performed  as shown by \citet{P08}.
Consider a photon path originating from a radius $r_{\rm I}$
that  has an angle
 $\theta_{v, {\rm I}}$ with respect to the fluid velocity (radial) direction
at the original position.
In this case,
the optical depth 
to reach  a  radius  $r_{\rm II}$
along the straight photon path can be expressed as  
%for the photon initially
%located at the radius $r_{\rm I}$ that 
%propagate in the direction having angle $\theta_{v, {\rm I}}$ with respect
%to the local velocity (radial) direction
%to reach radius $r_{\rm II}$  %angle $\theta_{v, {\rm II}}$ 
% can be expressed as
\begin{eqnarray}
\label{analtau}
\Delta \tau_{\rm I-II} &=&
 2 \Gamma^2 r_{\rm ph} 
\left[ \frac{\theta_{v,{\rm II}}}{r_{\rm II}{\rm sin}\theta_{v,{\rm II}}}
- \frac{\theta_{v, {\rm I}}}{r_{\rm I}{\rm sin}\theta_{v,{\rm I}}}
 + \beta \left(\frac{1}{r_{\rm I}} - \frac{1}{r_{\rm II}} \right) \right] \nonumber \\
&=& 
 2 \Gamma^2 \frac{r_{\rm ph}}{r_{\rm I}}
\left[ \frac{\theta_{v,{\rm II}} - \theta_{v, {\rm I}}}
{{\rm sin}\theta_{v,{\rm I}}}
 + \beta   
\left(1 - 
      \frac{{\rm cos}\theta_{v,{\rm II}}}{{\rm cos}\theta_{v,{\rm I}}} \right) \right] ,
\end{eqnarray}
%where $r_{\rm I}$ ($r_{\rm II}$) and $\theta_{v,{\rm I}}$
% ($\theta_{v,{\rm II}}$)
%are the initial (final) radius of the photon and its angle
%with respect to the local radial velocity direction
%\citep[for detail, see][]{P08}.
where $\theta_{v,{\rm II}}$ is the  photon angle
at the final position ($r=r_{\rm II}$).
In the second equality, we have used the relation
$r_{\rm I}{\rm sin} \theta_{v, {\rm I}} 
= r_{\rm II}{\rm sin} \theta_{v,{\rm II}}$
that holds for an arbitrary straight line.
Hence,
 in this case, we determine the corresponding propagation length
$\Delta l_{\rm I-II} = r_{\rm II}{\rm cos}\theta_{v, {\rm II}} - 
r_{\rm I}{\rm cos}\theta_{v, {\rm I}}$
from the drawn optical depth
 by solving equations (\ref{analtau}). 
(Note that equation (\ref{analtau}) is solved separately
 in the spine and sheath region, since
 the values of $r_{\rm ph}$ and $\Gamma$ $(\beta)$
 are different in each region).

On the other hand, when the photons have 
higher energies ($h \nu_{\rm cmf} >   100~{\rm keV}$)
% 0.1 m_e c^2$)
or are located below the saturation radius ($r<r_{\rm s}$),
the integration is solved numerically.
%For a given mesh,
%the interval of the numerical integration $dl$ is
%determined by the
%distance of the intersection points of the photon path
%(straight line) with the mesh interfaces.
In this case, 
we divide
the calculation
 region into a mesh in spherical coordinates ($r$,$\theta$,$\phi$).
Then the integration is done
by assuming that
the physical quantities (velocity and number density)
%along the photons path $\Delta l$ %which are contained
 within the individual mesh are uniform 
and  have the values corresponding to those
at the position of the mesh center.
%(The physical quantities are updated
%to the corresponding neigboring mesh
%when the photons crosses the mesh interface).
%
We adopt 500 grid points which are logarithmically spaced
for the mesh in the $r$ coordinate
between the inner boundary $r_{\rm in}$ and
outer boundary $r_{\rm out}$.
As for the mesh in the $\theta$ coordinate, 
we adopt 800 uniformly spaced grid points 
in the range $\theta \leq \theta_1$.
% for each intervals corresponding to
% the spine region ($0\leq \theta \leq \theta_0$) 
%and sheath region ($\theta_0 < \theta \theta_1$).
1600 uniformly spaced grid points
are 
adopted for the $\phi$ coordinate  ($0\leq \phi < 2 \pi$).
It is noted that the resolution of the grid is 
sufficiently high to 
reproduce the result
obtained by the analytical solution given by
equation (\ref{analtau}) 
(corresponding to infinite resolution) in cases when
$r \geq r_{\rm s}$ and $h \nu_{\rm cmf} \leq   100~{\rm keV}$
%0.1 m_e c^2$ 
are satisfied.

Given the position for the scattering  from the above procedure, 
the four-momentum (the energy and propagation direction) of a photon
after the scattering is determined
based on the differential cross section for 
Thomson and Compton scattering.
%differential Compton cross section.
In our code, 
the scattering process is evaluated in the
rest frame of the fluid (electron).
%using the full Klein-Nishina cross section.
First, the four-momentum of the photon before the scattering
is Lorentz transformed into the fluid rest frame.
For the photons that satisfy $h \nu_{\rm cmf} \leq   100~{\rm keV}$,
 the differential Thompson cross section is used, while
 the differential Compton cross section is used at higher 
 energies ($h \nu_{\rm cmf} >   100~{\rm keV}$).
%From the differential cross section, % given by the Klein-Nishina formula, 
The scattering angle or equivalently
the propagation direction of the outgoing photon
 in the fluid rest frame is drawn from the differential cross sections.
%the scattering angle of the outgoing photon in the fluid rest frame is drawn,
%and the propagation direction of the outgoing photon is determined.
Regarding the energy of the outgoing photons,
we assume that  it is conserved
 before and after the scattering %is conserved
(elastic scattering)
in the case of Thomson scattering
 ($h \nu_{\rm cmf} \leq   100~{\rm keV}$).
On the other hand, in the case of Compton scattering 
($h \nu_{\rm cmf} >   100~{\rm keV}$), energy loss due to the 
 recoil effect is properly taken into account.
The code then Lorentz transforms the outgoing photon four-momentum 
 back into the laboratory frame.

The above procedure is repeated
until all the injected photons
reach the boundary of the simulation grids.
%More detail on the treatment of the scattering process
%and test calculations are described in \S\ref{MCcodetest}.

\section{RESULTS}
\label{result}
In this section, we show the obtained photon spectra based
on the model described in the previous section.
We inject $N=2\times 10^8$ photon packets
in each calculation.
%Since the most of the photons are emitted within 
%a cone with half-opening angle $\sim \Gamma^{-1}$
%in the direction
%of the fluid velosity (radial direction) due to
%the relativisitic beaming effect, the emissions are highly anisotropic
%and varies largely depending on the angle between
%direction to the observer and jet axis, $\theta_{\rm obs}$. 
%In evaluating the spectrum for observer
%located at direction of $\theta_{\rm obs}$,
%we record all pariticles which have reached the outer 
%boundary that are propagating in direction 
%within a cone of half-opening angle $(2\Gamma)^{-1}$
%and calculate isotropic equivalent luminosity.
%spine and sheath are fixed at  $\theta_0 = 0.5^{\circ}$
In all cases, 
we employ a fixed value of
 $\theta_1 = 1^{\circ}$ for the half opening-angle of the jet,
which is smaller than that of the typically observed values.
It is emphasized, however,
that
the resulting spectra do not vary for wider jets (larger $\theta_1$)
as long as the 
observer angle stays in the range
 $\theta_{\rm obs} \lesssim \theta_1 - \Gamma^{-1}$.
This is simply because the emission from  regions
located at a angle $|\theta - \theta_{\rm obs}| \gtrsim \Gamma^{-1}$
is negligible due to the relativistic beaming effect.
Therefore, merely to reduce  computational cost,
we adopt the relatively small value in this study.
%It is noted, however, that
%almost same results are obtained in the case of
%the wider jet which have same structure considered
%in the present study at the center if 
%the angle between the direction of the  observer 
%and the jet-axis is 
%$\theta_{\rm obs} \lesssim \theta_1 - \Gamma^{-1}$,
%since the emissions from the region which  angle with 
%respect to observer direction larger than $1/\Gamma$ are
%negligible due the the relativistic beaming effect.
We use $r_{\rm i}=10^8~{\rm cm}$ 
for the initial radius of the fireball in all cases.

\subsection{Uniform (Non-Stratified) Jet}
\label{uniform}

Before
we look at a stratified jet, 
we first present results for a one-component uniform jet
that does not have structures in the $\theta$ direction
 ($\theta_0 = \theta_1 = 1^{\circ}$).
In this case,  a thermal spectrum is expected,
in contrast to a stratified jet as we will discuss later.
% unlike in the case where structure is present as we discuss later.
The isotropic equivalent kinetic luminosity and the dimensionless entropy
(terminal Lorentz factor)
are set to be
$L_0=10^{53}{\rm erg/s}$ and $\eta_0=400$, respectively.
As described in the previous section, %in this case,
we inject the photons at a radius of
$r_{\rm inj}= 4\times 10^{10}\eta_{0,400}r_{\rm i,8}~{\rm cm}$
with intensity given by a blackbody of
 temperature  
$k_{\rm B} T'_{\rm inj} = 
1.7r_{\rm i, 8}^{-1/2}\eta_{0, 400}^{-1}L_{0,53}^{1/4}~{\rm keV}$
 (see \S\ref{model} for detail), where
 $\eta_{0,400}=\eta_0/400$, $L_{0,53}=L_0/10^{53}~{\rm erg/s}$ and
 $r_{\rm i,8}=r_{\rm i}/10^{8}~{\rm cm}$. 
%and $k_{\rm B}$ is the Boltzmann constant.
The corresponding optical depth at the injection radius 
is $\tau(r_{\rm inj}) \sim 23$.
The results are insensitive to
the value of $r_{\rm inj}$ as
long as $\tau \gg 1$ is satisfied as noted in \S\ref{model}. 
%
%Regarding the subequent evolution of the injected photons, 
%Initially, when the optical depth is sufficiently high $\tau \gg 1$,
%photons are strongly coupled with the fluid via multiple scattering with
%the electrons and 
%advect outward with the fluid. 
The advected photons lose energy adiabatically due to the expansion of the flow
until the coupling with  matter becomes weak near the photosphere
%,%which is located at
($r_{\rm ph0} \simeq 9.2\times 10^{11}L_{0,53}\eta_{0,400}^{-3}~{\rm cm}$).
The expected temperature at the photosphere can be calculated
as  
$k_{\rm B} T'_{\rm ph0} = 
0.38 r_{\rm i,8}^{1/6} \eta_{0,400}^{5/3}L_{0,53}^{-5/12}~{\rm keV}$.
The observed peak energy of the photospheric emission is
expected to be $E_{\rm p} \sim 8 \eta_0 k_{\rm B} T'_{\rm ph0}
\sim 660 r_{\rm i,8}^{1/6} \eta_{0,400}^{8/3}L_{0,53}^{-5/12}~{\rm keV}$, 
%where $h$ is the Plank constant,
since the photon energy is boosted by a factor ${\cal D} \sim 2 \eta_0$ due
to the Doppler effect.
Also the luminosity of the emission can be estimated
as $L_{\rm p}\sim L_0(r_{\rm ph0}/r_{\rm s0})^{-2/3}\sim
    1.2\times10^{52} r_{\rm i,8}^{2/3} \eta_{0,400}^{8/3}L_{0,53}^{1/3}~{\rm erg/s}$.

%The corresponding 
%injection radius and the 
% photosphere are located at $r_{\rm inj}= 4\times 10^{10}\eta_{0,400}r_{\rm i,8}~{\rm cm}$ and
% $r_{\rm ph0} \simeq 9.2\time 10^{11}L_{0,53}\eta_{0,400}^{-3}~{\rm cm}$, where
% $\eta_{0,400}=\eta_0/400$, $L_{0,53}=L_0/10^{53}~{\rm erg/s}$ and
% $r_{\rm i,8}=r_{\rm i}/10^{8}~{\rm cm}$.
%The resultant spectrum is shown in Fig. \ref{nost}

\begin{figure}[ht]
\begin{center} 
\includegraphics[width=9cm]{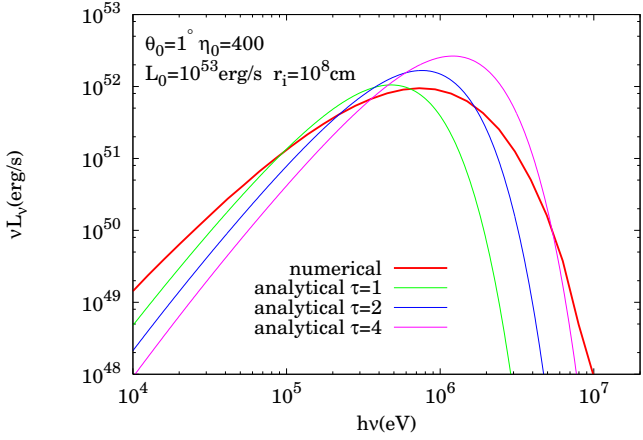}
\caption 
{Observed luminosity spectrum in the case of a uniform jet
 ($\theta_0 = \theta_1 = 1^{\circ}$)
  with
 parameters of the fireball
 given by $L_0=L_1=10^{53}~{\rm erg/s}$, $\eta_0=\eta_1=400$ and
 $r_{\rm i}=10^8~{\rm cm}$.
% The half opening angle of the jet is taken as $\theta_1=1^{\circ}$. 
 The spectrum corresponds to the case when the line of sight 
 to the observer is aligned to the jet axis
 ($\theta_{\rm obs}=0^{\circ}$).
 The red line shows the numerical result.
 The green, blue and purple lines show the
 analytical estimates of photospheric emission derived 
 from equation (\ref{photoana})
 for $r = r_{\rm ph}$, $r = r_{\rm ph}/2$ and $r = r_{\rm ph}/4$, respectively.} 
\label{nost}
\end{center}
\end{figure}

The  numerical result is displayed in Fig. \ref{nost} with a
red solid line. 
Here we assume that the observer is
 aligned to the jet axis ($\theta_{\rm obs}=0^{\circ}$).
It is noted, however, that the result does not change if  
the observer angle stays in the range
$\theta_{\rm obs} \lesssim \theta_1 - \eta_0^{-1} \sim 0.86^{\circ}$,
since the fluid properties within the beaming cone
($|\theta - \theta_{\rm obs}|\lesssim 0.14^{\circ}\eta_{0,400}^{-1}$)
do not change.
%axis-symmetric with respect to the line of sight 
%within the beaming cone in this case.
In the figure, we also show an analytical solution for the 
expected emission 
when the photons are in complete thermal equilibrium up to 
a radius $r$ 
which can be obtained as
\begin{eqnarray}
\label{photoana}
L_{\nu}(\nu) = 8\pi^2 r^2 \int^1_{{\rm cos}\theta_1} && 
              {\cal D}(\eta_0, \theta)^3 B_{\nu}(T'(r), \nu/{\cal D}(\eta_0,\theta)) \times \nonumber \\ 
	    &&  {\rm cos}\theta~{\rm d cos}\theta , 
\end{eqnarray}
where $\theta$ is angle between the line of sight and
fluid velocity (radial) direction.
%where $T_{\rm ph0} = 
%0.38 r_{\rm i,8}^{1/6} \eta_{0,400}^{5/3}L_{0,53}^{-5/12}~{\rm keV}$ 
%is the temperature at the photospheric radius.
The green, blue and purple solid lines
correspond to the solutions for $r=r_{\rm ph0}(\tau = 1)$,
 $r=r_{\rm ph0}/2(\tau = 2)$
 and 
$r = r_{\rm ph0} / 4(\tau = 4)$, respectively.
Regarding the case of $r=r_{\rm ph0}$,
the peak energy, $E_{\rm p}$, and
luminosity, $L_{\rm p} \sim \nu_{\rm p} L_{\nu}(\nu_{\rm p})$,
 where $\nu_{\rm p}=E_{\rm p}/h$, 
of the obtained spectrum
agree  well with 
the rough estimate given earlier.
%
%Note that the peak energy $E_{\rm p} \sim 500~{\rm keV}$ 
%is slightly higher than the earlier estimate.
%The small discrepancy is due to the fact that
%the Doppler factor ${\cal D}(\eta_0,\theta)$ varies
%from $\sim \eta_0$ to $\sim 2\eta_0$ in
%the range $\theta < 1/\Gamma$, where the majority of the
%emissions are produced,
%while in the earlier estimate was based on the assumption of
%${\cal D}\sim \eta_0$.  
%the obtained spectrum shows a peak at energy
%$E_{\rm p} \sim 500~{\rm keV}$ which
%is somewhat higher than the rough estimate given earlier. 
%The small discrepancy is due to the fact that
%the Doppler factor ${\cal D}(\eta_0,\theta)$ varies
%from $\sim \eta_0$ to $\sim 2\eta_0$ in
%the range $\theta < 1/\Gamma$, where the majority of the
%emissions are produced,
%while in the earlier estimate was based on the assumption of
%${\cal D}\sim \eta_0$. 
%It is also noted that the luminosity of the emission,
% $L_{\rm p} \sim [\nu_{\rm p} L_{\nu}(\nu_{\rm p}$,
% where $\nu_{\rm p}=E_{\rm p}/h$
% agrees quite well with a rough estimate given earlier.
%
As for the cases of $r=r_{\rm ph0}/2$ 
and  $r=r_{\rm ph0}/4$, 
both the peak energy and the 
luminosity are larger
by a factor $\sim 2^{2/3} \sim 1.6$ and 
 $\sim 4^{2/3} \sim 2.5$
than the case of $r=r_{\rm ph}$, respectively,
 since
the temperatures
at these radii are larger  by
the same factor. % ($T(r_{\rm ph0}/5)/T_{r_{\rm ph0}}=5^{2/3}$).
As shown in the figure,
the peak luminosity of the numerical result is in good agreement with 
that of the analytical estimate for  $r = r_{\rm ph}$.
On the other hand, the spectrum  extends up to higher energies and
the peak energy is close to that for $r=r_{\rm ph}/2$.
%rather than $r=r_{\rm ph}$.
This is due to
the fact that
the coupling between the photon and matter is not complete
near the photosphere $\tau \sim 1$ as shown in
the previous studies \citep{P08,  B11, BSV13}.
As a result,  photons which decouple with the matter
 at moderate optical depth
($\tau \lesssim 5$) are observed at  higher energies.

Regarding the shape of the spectrum,
while the emission is dominated by  
photons that escaped from the on-axis region ($\theta \lesssim \Gamma^{-1}$)
at energies
near the peak energy and above,  
the low energy part ($\nu \ll \nu_{\rm p}$)
is dominated by those 
from the off-axis region. % which have decoupled with the matter at $r \sim r_{\rm ph}/5$.
%Therefore, the resultant spectrum can be well reproduced by the 
%analytic estimate.
The off-axis component becomes prominent at 
low energies because the Doppler factor is
smaller which leads to a lower peak photon energy
 $\sim 4 {\cal D}(\eta_0,\theta)k_{\rm B}T'$.
Therefore, the low energy part of the spectrum can be 
expressed as a superposition of Blackbody spectra
from the off-axis region which have different peak energies
(multi-color Blackbody).
 %and can be well approximated by the analytic estimate.
As a result, 
the low energy slope of the spectra is somewhat softer than
that expected from the Rayleigh-Jeans part of a single blackbody 
($\nu L_{\nu} \propto \nu^3$) and
can be roughly approximated as $\nu L_{\nu} \propto \nu^{2.4}$.

\subsection{Stratified Jet}
\label{stratified}

%{\bf
%\subsubsection{Infinitesimal boundary layer}
%}

 Here we show the results for a two-component stratified jet.
 In all cases, the half opening angle of the spine is fixed at
 $\theta_0 = 0.5^{\circ}$.

 As mentioned in \S\ref{uniform}, when a uniform jet is assumed, spectra
 tend to be thermal-like with slight modifications from 
 blackbodies originating from
 a sphere with radius $r=r_{\rm ph0}/5$.
 On the other hand, the appearance of the spectrum
 can deviate significantly
 from a thermal one  when a strong velocity shear is 
 present in the outflow, since photons which cross
 the shear flow multiple times can gain energy through a
 Fermi-like acceleration mechanism.
 This can be understood as follows.
 Under the assumption of elastic scattering, 
% in every scattering,
 the energy gain of photons in a single scattering event
% the change in the  photon energy in each scattering
 can be expressed as
%$\nu_{\rm sc}/\nu_{\rm in} = 
% (1-\beta {\rm cos} \theta_{\rm in})/(1-\beta {\rm cos} \theta_{\rm sc})$,
 \begin{eqnarray}
  \label{rate}
  \frac{\nu_{\rm sc}}{\nu_{\rm in}} = 
  \frac{1-\beta {\rm cos} \theta_{\rm in}}{1-\beta {\rm cos} \theta_{\rm sc}},
 \end{eqnarray}
 where $\nu_{\rm in}$ ($\nu_{\rm sc}$) and
  $\theta_{\rm in}$ ($\theta_{\rm sc}$) 
 are the frequency and
 angle between the the fluid velocity and photon propagation direction
 before (after) the scattering, respectively.
 Hence, if $\theta_{\rm sc} < \theta_{\rm in}$,
 photons  gain energy and vice versa.
 Photons which have crossed the 
 boundary layer from the sheath to the spine region
 tend to gain energy when they are scattered there
 (upscatter).
 This is simply because
 the photons within the sheath region 
 tend to have larger angle between their propagation direction
 and  fluid velocity than those in the spine region. 
% since the photons advecting with matter
% in the sheath region
% tend to have larger angle between the photon direction and 
% fluid velocity than those in the spine region.
% This is simply because relativistic beaming effect 
% is stronger in the spine region due to the larger
% bulk Lorentz factor.
 On the other hand, photon which have crossed 
 the boundary layer from the spine to the sheath region
 tend to lose energy (downscatter) due to the same reason.
 Consequently, some fraction of 
 photons which %experience multiple crossing of the boundary layer
 cross the boundary layer multiple times can gain energy,
 since the 
 energy gain by the upscattering %in the spine region 
 overcome the downscattering in average. %in the sheath region on average.
 This mechanism can give rise to a non-thermal spectrum
 at the high frequencies.

 To obtain a rough estimation of the
 average energy gain and loss rate ($\nu_{\rm sc}/\nu_{\rm in}$) for 
 each process, let us approximate the radially expanding
 spine and sheath regions as a plane parallel flow.  
 Under the above consideration, the typical angle between 
 the photon propagation direction and the fluid velocity direction 
 for the photons in the spine (sheath) region can be roughly estimated as
 $\langle{\theta_v}\rangle_{0} \sim \Gamma_0^{-1}$
 ($\langle{\theta_v}\rangle_{1} \sim \Gamma_1^{-1}$).
 Since the angle $\theta_v$ is conserved along the photon's path in the case of a plane parallel flow,
 the typical energy gain rate by the upscattering
 in the spine region can be evaluated by substituting
 $\theta_{\rm in}=\langle{\theta_v} \rangle_1 \sim \Gamma_1^{-1}$ and
 $\theta_{\rm sc}=\langle \theta_v \rangle_0 \sim \Gamma_0^{-1}$ 
 in equation (\ref{rate})
 and is given as
 \begin{eqnarray}
 \label{up}
   \left\langle \frac{\nu_{\rm sc}}{\nu_{\rm in}} \right\rangle_{\rm up} \sim
   \frac{1-\beta_0 {\rm cos} \langle{\theta_v}\rangle_1}
	{1-\beta_0 {\rm cos} \langle{\theta_v}\rangle_0}
   \sim \frac{1}{2}\left\{1+ \left(\frac{\Gamma_0}{\Gamma_1}\right)^2 \right\}.
 \end{eqnarray}
 Similarly, the typical energy loss rate by the downscattering
 in the sheath region is given as
 \begin{eqnarray}
 \label{down}
   \left\langle \frac{\nu_{\rm sc}}{\nu_{\rm in}} \right\rangle_{\rm down} \sim
   \frac{1-\beta_1 {\rm cos} \langle{\theta_v}\rangle_0}
	{1-\beta_1 {\rm cos} \langle{\theta_v}\rangle_1}
   \sim \frac{1}{2}\left\{1+ \left(\frac{\Gamma_1}{\Gamma_0}\right)^2 \right\}.
 \end{eqnarray}
 From the above equations, it is clear that 
 the energy gain by the upscattering %in the spine region 
 overcomes the energy loss by  the downscattering  
 ($\langle \nu_{\rm sc}/\nu_{\rm in} \rangle_{\rm up} \langle \nu_{\rm sc}/\nu_{\rm in} \rangle_{\rm down} \sim
 (1/4) [2 + (\Gamma_0/\Gamma_1)^2 + (\Gamma_1 / \Gamma_0)^2] > 1$).
% Consequently, 
% fraction of 
% photons which %experience multiple crossing of the boundary layer
% cross the boundary layer multiple times tend to gain energy.
% This mechanism can give rise to a non-thermal spectrum
% at the high frequencies.
 It is also clear that the efficiency of the acceleration
 per each cycle of crossing 
 $\langle \nu_{\rm sc}/\nu_{\rm in} \rangle_{\rm up} \langle \nu_{\rm sc}/\nu_{\rm in} \rangle_{\rm down} \sim
 (1/4)[2 + (\Gamma_0/\Gamma_1)^2 + (\Gamma_1 / \Gamma_0)^2]$
 is controlled by the ratio between the bulk Lorentz factor
 of the two regions
  $\Gamma_0 / \Gamma_1$ and 
 increases as the ratio becomes larger.
% Note that the acceleration efficiency
% is somewhat smaller in the radially expanding flow, since
% the angle  $\theta_v$ decreases motononically along the straight photon
 %which in turn leads to reduction of energy gain (loss) rate.
%
 It is worth noting that,
 while the {\it average} value of the 
 energy ratio $\nu_{\rm sc}/\nu_{\rm in}$ roughly obeys the above equations,
 the dispersion  around the average value is large, since
 it depends quite sensitively on the scattering angles
  ($\theta_{\rm in}$ 
 and $\theta_{\rm sc}$; see Eq. (\ref{rate})).
 When a photon from the sheath region that has  
 an angle $\theta_{\rm in}=f_1 \Gamma_1^{-1}$ is scattered
 in the spine region with an angle
 $\theta_{\rm sc} = f_0 \Gamma_0^{-1}$, the 
 energy gain by the scattering can be written as
 $\nu_{\rm sc}/\nu_{\rm in} 
\sim (1+f_0^2)^{-1}[1+f_1^2(\Gamma_0/\Gamma_1)^2]$.
%
% For an incoming photon from the sheath region 
% that has an angle $\theta_{\rm in}=f_1 \Gamma_1^{-1}$ that 
% is scattered at the spine region with an angle
% $\theta_{\rm sc} = f_0 \Gamma_0^{-1}$,
% the energy ratio can be written as 
% $\nu_{\rm sc}/\nu_{\rm in}\sim (1+f_0^2)^{-1}[1+f_1^2(\Gamma_0/\Gamma_1)^2]$.
%
%
 It is clear from the above equation that 
 a small change in the scattering angles %from the typical values
 ($\theta_{\rm in}=\langle{\theta_v}\rangle_{0} \sim \Gamma_0^{-1}$ and
  $\theta_{\rm sc}=\langle{\theta_v}\rangle_{1} \sim \Gamma_1^{-1}$)
 leads to a quite large change in the energy ratio.
 For example,
 in the case of $f_0=0$ and  $f_1=2$, the
 energy ratio due to the upscattering is 
 larger than the typical value by a factor of 
 $(\nu_{\rm sc}/\nu_{\rm in})
 (\langle \nu_{\rm sc}/\nu_{\rm in} \rangle_{\rm up})^{-1}
 \sim 2[1+4(\Gamma_0/\Gamma_1)^2][1+(\Gamma_0/\Gamma_1)^2]^{-1}\sim 8$.
% photon that have an angle $\theta_{\rm in}$
% larger than the typical value by a factor of $\sim 2$ ($f_1=2$),
%
% the energy ratio is larger the typical energy gain given in Eq. (\ref{up})
% roughly by a factor of $\sim f_1^2$.
%
% For 
% an incoming photon from the sheath region 
% that have an angle $\theta_{\rm in}=f_1 \Gamma_1^{-1}$ that 
% is scattered at the spine region with the typical angle
 %$\theta_{\rm sc} = \langle{\theta_v}\rangle_{0} \sim \Gamma_0^{-1}$,
% the energy ratio can be written as
% $\nu_{\rm sc}/\nu_{\rm in}\sim 1/2[1+f_1^2(\Gamma_0/\Gamma_1)^2]$..
%
% Since non-negligible fraction of photons can have angles with
% factor of few larger than 
% In principle, 
% the energy ratio can range from
% $(4\Gamma^{2})^{-1}$ up to $4\Gamma^{2}$
% depending the the scattering angles (see Eq. (\ref{rate})).
% 
% Note that the energy gain (loss)
% in the radially expanding flow is smaller 
% than that given by the equation, 
%  since the angle 
% $\theta_v$ decreases monotonically along the straight photon.
%
Note also that, once the photon 
energy (evaluated in the electron rest frame)
approaches close to the electron rest mass energy
$h \nu_{\rm cmf} \sim m_e c^2$, where $m_e$ is the electron rest mass,  
the scattering can no longer be approximated as elastic, since
recoil effect becomes non-negligible (Klein-Nishina effect).
In this case, the acceleration efficiency is significantly  reduced.
%
% Therefore, the efficiency of acceleration is % somewhat
% reduced  in the actual calculations.

% \begin{eqnarray}
%   \nu_{\rm sc}
% \end{equnarray}

% Since the Lorentz factor is larger in
% the spine region,
% the relativistic beaming effect is stronger than that 
% in the sheath region.
% Therefore,
% photons 
% tend to have smaller angle between the photon direction and 
% fluid velosity than those in the spine region
% For simplisity, let us approximate the 
% spine-sheath structure (Fig. \ref{modelpic})
% as a plane pararell flow propagating in a same direction
% with different Lorentz factor rather than radial outflow.
% In this case, the angle between the
% direction of fluid velosity and photon $\theta_v$, 
% does not vary along the photon path.
% (The value of $\theta_v$ vary along the ray in the case of the radial flow).
% If the photons are isotropic in the fluid rest frame,
% the expected value of 

% Photons 
% which have crossed the boudary layer from the sheath region to the spine 
% gain energy on average, since  

\begin{figure}[ht]
\begin{center} 
\includegraphics[width=9cm]{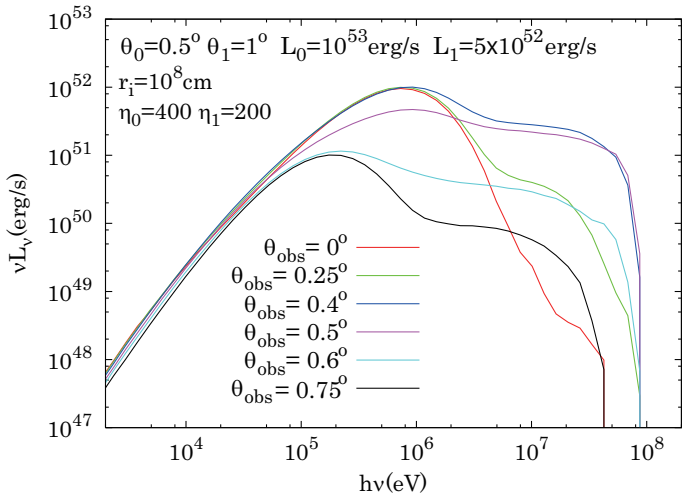}
\caption 
{Observed luminosity spectrum in the case of spine-sheath jet
 in which the spine jet with half opening angle of $\theta_0=0.5^{\circ}$
 is embedded in a wider sheath outflow with half opening angle of
 $\theta_1=1^{\circ}$.
 The employed
 values for dimensionless entropy (terminal Lorentz factor)
 and kinetic luminosity are chosen
 as $\eta_0=400$ and  $L_0=10^{53}~{\rm erg/s}$ for the spine and 
  $\eta_1=200$ and  $L_1=(\eta_1/\eta_0)L_0=5\times10^{52}~{\rm erg/s}$ for the sheath, 
 respectively.
 The initial radius of fireball is chosen as  
 $r_{\rm i}=10^8~{\rm cm}$ in both regions.
 The various lines show the cases where the
 observer angle with respect to the jet axis is
 $\theta_{\rm obs}=0^{\circ}$ (red), $0.25^{\circ}$ (green),  $0.4^{\circ}$ (blue), 
 $0.5^{\circ}$ (purple),  $0.6^{\circ}$ (light blue) and  $0.75^{\circ}$ (black).
}  
\label{G400st}
\end{center}
\end{figure}

% When the structure in the fluid is considered, the appearance
% of the emissions become quite different since
% a fraction of photons can gain energy by .
 In Fig. \ref{G400st}, we display the obtained result
 for the case of a stratified jet with
% when values of
% the dimensionless entropy (terminal Lorentz factor)
% and kinetic luminosity are imposed as 
 $\eta_0=400$ and $L_0=10^{53}~{\rm erg/s}$ for
 the spine and 
 $\eta_1=200$ and  $L_1= (\eta_1/\eta_0)L_0 = 5\times10^{52}~{\rm erg/s}$
 for the sheath. %, respectively.
 As mentioned in \S\ref{model}, the injection radius
 is set at a position where a velocity shear between the two regions
 develops ($r_{\rm inj}=r_{\rm s1}$). 
 The corresponding optical depth is
 $\tau(r_{\rm inj}) \sim 100$ for the spine
 and $\tau(r_{\rm inj}) \sim 180$ for the sheath. 
% $\tau(r_{\rm inj}) \sim 104$ for the spine 
% and $\tau \sim 184$ for the sheath.
%
 The various lines in the figure show the cases for the
 observer angle with respect to the jet axis 
 being 
 $\theta_{\rm obs}=0^{\circ}$ (red), $0.25^{\circ}$ (green), 
 $0.4^{\circ}$ (blue), 
 $0.5^{\circ}$ (purple), 
 $0.6^{\circ}$ (light blue)
 and  $0.75^{\circ}$ (black).
 As we can see, 
 the spectrum varies quite sensitively with the observer angle.
 The spectrum for $\theta_{\rm obs}=0^{\rm \circ}$ is 
 thermal-like and 
  nearly identical to that obtained in the case of a uniform jet 
 (Fig. \ref{nost}).
 The reason for this is simple.
 Since most of the scattered photons propagate in a direction
 within a cone of  half opening angle
 $\sim 1/\Gamma \sim 0.14^{\circ}(\Gamma/400)^{-1}$,
 the majority of the observed 
 photons are  from a region of $\theta \lesssim 0.14^{\circ}$.
 Hence, only a  small fraction of photons from the sheath region
 and the boundary
 ($\theta \geq \theta_0 = 0.5^{\circ}$) can reach the observer,
so that the spectrum does not deviate  largely
 from the case of uniform jet.
 On the other hand, if the observer angle is larger,
 photons from the sheath and boundary layer become observable.
 As a result, a 
 non-thermal component appears above the 
 peak energy of the thermal spectrum
 due to the photon acceleration in the boundary layer.
 The non-thermal component is hardest 
 when the observer angle is aligned to the 
 boundary layer $\theta_{\rm obs}=\theta_0 = 0.5^{\circ}$
 and becomes softer as the deviation between 
 $\theta_{\rm obs}$ and $\theta_0$ becomes larger, simply
 because the boundary layer corresponds to the site of 
 photon acceleration.
 As mentioned earlier, 
 the photon acceleration becomes inefficient when the
 photon energy becomes large enough so that the
 recoil
 of electrons cannot be neglected (Klein-Nishina effect).
 Hence, in all cases, the spectrum does not extend up to energies higher than
 $h\nu \sim \Gamma_0 m_e c^2 \sim 200(\Gamma_0/400)~{\rm MeV}$.

 Note also that the peak energy and 
 the luminosity of the thermal component
 differs enormously 
 for $\theta_{\rm obs} > \theta_0$ and
 for $\theta_{\rm obs} < \theta_0$, due 
 to the differences in the assumed parameters
 in the spine and sheath regions.
 For an observer at $\theta_{\rm obs} < \theta_0$, the 
 thermal component is determined mainly by 
 photons which have propagated through the spine region.
 Therefore,
 the observed spectrum is
 nearly identical to the case of the uniform jet 
 considered above in which a same set of parameters
 ($\eta_0$, $L_0$ and $r_{\rm i}$) is assumed.
 On the other hand,
 for an observer at $\theta_{\rm obs} > \theta_0$,
 photons which have propagated through the sheath region 
 dominate the thermal component.
 Accordingly, the peak energy and luminosity are lower
 by a factor $\sim (\eta_0/\eta_1)^{8/3}(L_0/L_1)^{-5/12}\sim 4.7$ and
 $\sim (\eta_0/\eta_1)^{8/3}(L_0/L_1)^{1/3}\sim 8$, respectively.

% The photon acceleration becomes inefficient when 
% the photon

\begin{figure}[ht]
\begin{center} 
\includegraphics[width=9cm]{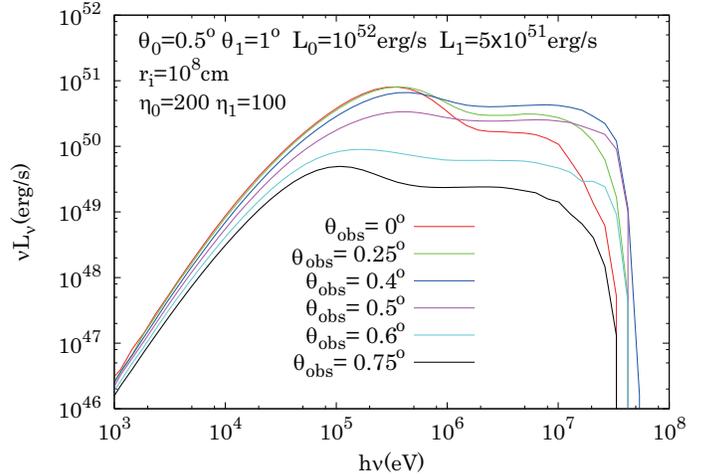}
\caption 
{Same as Fig.\ref{G400st}, but for
  $\eta_0=200$,
  $\eta_1=100$,
  $L_0=10^{52}~{\rm erg/s}$ and   $L_0=5\times10^{51}~{\rm erg/s}$.}  
\label{G200st}
\end{center}
\end{figure}

 In Fig. \ref{G200st}, we display the results obtained for
 $\eta_0=200$ and  $L_0=10^{52}~{\rm erg/s}$  for the spine and 
 $\eta_1=100$ and  $L_0=5\times10^{51}~{\rm erg/s}$ for the sheath.
 That is, the terminal Lorentz factor of the 
 outflow is smaller by a factor of $2$ for both the spine and the sheath
 than those assumed in the previous case.
 The optical depths at the injection radius are
 $\tau(r_{\rm inj}) \sim 170$ for the spine and 
 $\tau(r_{\rm inj}) \sim 290$ for the sheath. 
 %$\tau \sim 166$ for the spine and 
 %$\tau \sim 294$ for the sheath.
% 
 As in the previous case,
 the non-thermal component
 is hardest when %$\theta_{\rm obs}=\theta_0$
 $\theta_{\rm obs}=\theta_0 = 0.5^{\circ}$
 and becomes softer as the deviation between 
 $\theta_{\rm obs}$ and $\theta_0$ becomes larger.
 However, the major difference with the previous case
 is that the observer dependence of the hardness 
 is weaker.
 For example, as shown in Fig. \ref{G200st},
the  non-thermal component can be 
 prominent even for $\theta_{\rm obs}=0^{\circ}$.
 This is  because
 beaming effect is weaker than the previous case
 due to the smaller values of the Lorentz factor,
 so that
 the photons can spread out in wider angles.

 The dependence of the spectrum on the 
 difference between
 the dimensionless entropies (terminal Lorentz factor) of the spine
 and sheath is displayed 
 in Fig. \ref{G400compare}.
 Each panel
 corresponds to the result for the observer angle 
 fixed at
 $\theta_{\rm obs} = 0.25^{\circ}$ (top panel), 
 $\theta_{\rm obs} = 0.4^{\circ}$ (middle left), 
 $\theta_{\rm obs} = 0.5^{\circ}$ (middle right), 
 $\theta_{\rm obs} = 0.6^{\circ}$ (bottom left) and
 $\theta_{\rm obs} = 0.75^{\circ}$ (bottom right). 
%
% In all cases, the observer angle
% fixed as  $\theta_{\rm obs} = 0.4^{\circ}$, and
 In all cases, the
 dimensionless entropy
 and kinetic luminosity
 of the 
 spine are chosen to be
 $\eta_0 = 400$ and 
 $L_0=10^{53}~{\rm erg/s}$, respectively.
 The red line shows the case for a uniform jet,
 while
 the green, blue, purple, light blue and black lines show
 the cases
 for a sheath with dimensionless entropies of 
 $\eta_1=300$, $\eta_1=250$,  $\eta_1=200$,  $\eta_1=150$
 and  $\eta_1=100$, respectively.   
 The kinetic luminosity of the sheath is
 determined by $L_1=(\eta_0/\eta_1)L_0$.

 As mentioned earlier,
 the peak energy and  luminosity of the thermal-component
 depend on the dimensionless entropy and the kinetic luminosity as
 $E_{\rm p} \propto \eta^{8/3}L^{-5/12}$
 and luminosity $L_{\rm p}\propto \eta^{8/3}L^{1/3}$.
 Hence, 
 for observer mainly seeing the photons from the sheath
 region ($\theta_{\rm obs}>\theta_0$),
 these values show considerable decrease  
 in  models assuming smaller  $\eta_1$
 as is seen in the figure.
% Note that the
 %significant decrease in the peak energy and luminosity
%
% Regarding the non-thermal component,
% As is seen in the figure,
 The non-thermal component becomes significant
 as  $\eta_1$ becomes smaller.
 This tendency is due to the following reasons.
 One is simply because 
 the bulk Lorentz factor of the sheath becomes smaller for
 smaller $\eta_1$.
 As a result, the ratio between the bulk Lorentz factor 
 of the spine and sheath 
 $\Gamma_0/\Gamma_1$
 becomes larger which in turn leads
 to an increase in the energy gain per each crossing
 as explained earlier in this section.
% As a result, the photons in the sheath region
% become less concentrated in the radial direction since
% the relativistic beaming effect becomes weaker.
% relativistic beaming effect of the photon
% in the sheath becomes smaller. %, while that in the spine remains unchanged.
% This increases the
% average difference between $\theta_{\rm in}$ and 
% $\theta_{\rm sc}$
% for photons crossing the boundary layer 
% which in turn leads to an increase in the average energy gain
% per each crossing.
%
 In addition,
 the wider spreading of the photons propagating in the sheath region 
 due to the increase in the beaming angle $\sim \Gamma_1^{-1}$
 increases the chance for the photons to cross
 the boundary layer from the sheath to the spine region.
% which in turn
% increases the population of the photons to be accelerated.
%
 Another reason is that, for smaller value of $\eta_1$, 
 the radius where the velocity shear begins to develop 
 $r_{\rm s1}=\eta_1 r_{\rm i}$
 becomes smaller.
 This also leads to an increase in the 
 probability for the photons to be accelerated,
 since the optical depth of the acceleration region ($r>r_{\rm s1}$) 
 increases (see equation (\ref{tau})).
% The corresponding optical depth at the 
% spine region increases simply because $r_{\rm s1}=\eta_1 r_{\rm i}$
% decreases.
% In the sheath region, in addition to 
% the decrease in the $r_{\rm s1}$, 
% the decrease in $\eta_1$ increases
% the corresponding optical depth since 
% opacity becomes larger (see equation (\ref{tau})).
% 

\subsection{Relation Between  Photon Energy and Number of Crossings}
\label{crdiss}

To demonstrate that  photons 
 accelerated via the multiple crossing of the spine and sheath boundary layer
are indeed the origin of the non-thermal component,  %the photon acceleration,
we analyzed the relation
between the photon energy and the number 
of crossings that the corresponding photons have experienced.
%in Fig.
%
Here the number of crossings, $n_{\rm cr}$, is defined as the
total number of events that the photon 
has crossed the boundary layer
 (either from spine to sheath or sheath to spine) before it reaches
the outer boundary $r_{\rm out}$. 
%Note that we do not count the event 
%of crossings, 
%if the photon reach the outer boundary without being scattered after
%the crossing occured, since it does not 
%
%We evaluate
%The average number of crossing for a given photon energy, 
% $\langle n_{\rm cr} \rangle_{n_{\rm cr}}$,
%is simply calculated by 
%adding up the number of event for all the photons at the energy 
%and then deviding it by the number of the corresponding photons.
%

%

\begin{figure*}[ht]
\begin{center}

%\onecolumn

\includegraphics[width=16cm]{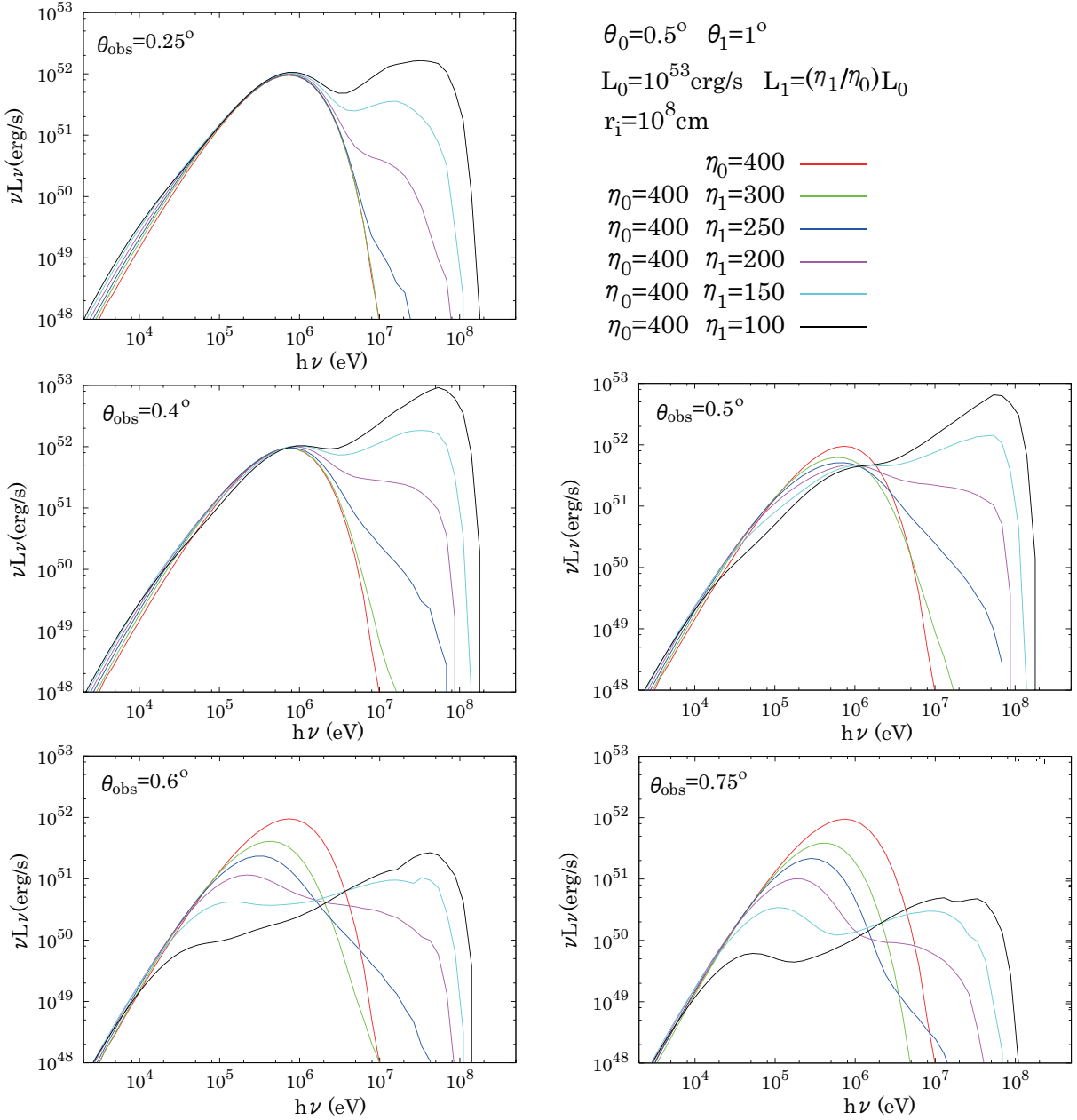}
\caption 
{Observed luminosity spectrum for a uniform jet and a
 spine-sheath jet
 in which the spine jet with half opening angle of $\theta_0=0.5^{\circ}$
 is embedded in a wider sheath outflow with half opening angle of
 $\theta_1=1^{\circ}$.
 Each panel shows a different observer angle fixed
 at $\theta_{\rm obs} = 0.25^{\circ}$ (top panel),
 $\theta_{\rm obs} = 0.4^{\circ}$ (middle left), 
$\theta_{\rm obs} = 0.5^{\circ}$ (middle right),
$\theta_{\rm obs} = 0.6^{\circ}$ (bottom left) and
$\theta_{\rm obs} = 0.75^{\circ}$ (bottom right).
 The employed
 values for dimensionless entropy (terminal Lorentz factor)
 and kinetic luminosity are chosen
 as $\eta_0=400$ and  $L_0=10^{53}~{\rm erg/s}$ for the spine.
 The red line shows the case of a uniform jet, while
 the green, blue, purple, light blue and black show
 the cases for the dimensionless entropy of the sheath
 given by $\eta_1=300$, $\eta_1=250$,  $\eta_1=200$,  $\eta_1=150$
 and  $\eta_1=100$, respectively.   
 In each case, the kinetic luminosity of the sheath is
 given by $L_1=(\eta_0/\eta_1)L_0$.
} 
\label{G400compare}   
\end{center}
\end{figure*}

\begin{figure}[ht]
\begin{center} 
\includegraphics[width=9cm]{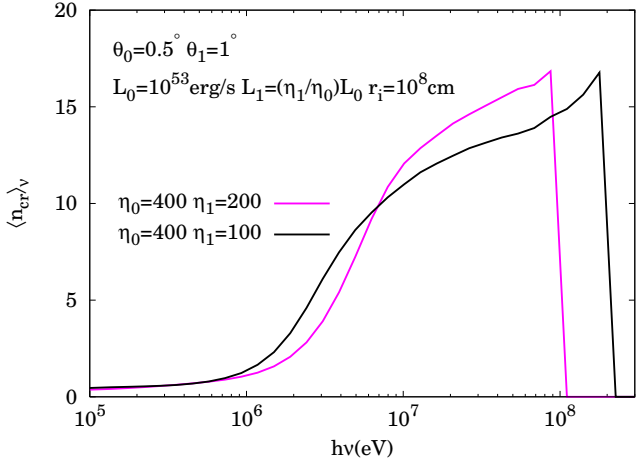}
\caption 
{
Average number of crossing events of the 
spine-sheath boundary layer
% $\langle n_{\rm cr} \rangle_{\nu}$
 as a function of observed photon energy for a jet
 in which the spine region with half opening angle of $\theta_0=0.5^{\circ}$
 is embedded in a wider sheath outflow with half opening angle of
 $\theta_1=1^{\circ}$.
 The employed
 values for dimensionless entropy (terminal Lorentz factor)
 and kinetic luminosity are chosen
 as $\eta_0=400$ and  $L_0=10^{53}~{\rm erg/s}$ for the spine.
 The purple and black lines show the cases of the dimensionless entropy of the sheath
 given by   $\eta_1=200$  and  $\eta_1=100$, respectively.
 In each case, the kinetic luminosity of the sheath is
 given by $L_1=(\eta_0/\eta_1)L_0$.
}
\label{CRave}
\end{center}
\end{figure}

\begin{figure}[ht]
\begin{center} 
\includegraphics[width=9cm]{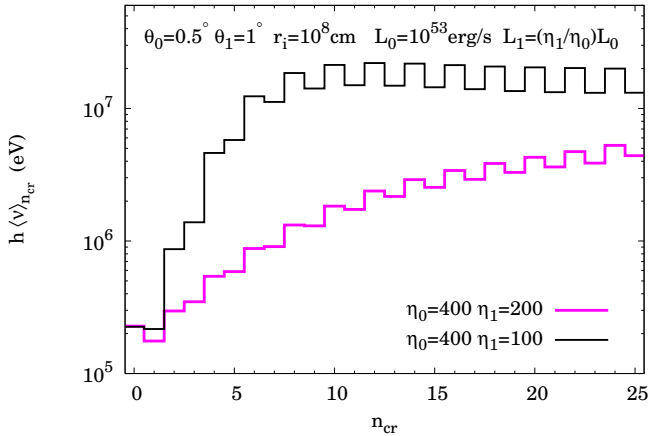}
\caption 
{
Average energy of the photons 
% $\langle \nu \rangle_{n_{\rm cr}}$
 as a function of the number of 
 crossings of the spine-sheath boundary layer $n_{\rm cr}$
for a jet
 in which the spine region with half opening angle of $\theta_0=0.5^{\circ}$
 is embedded in a wider sheath outflow with half opening angle of
 $\theta_1=1^{\circ}$.
 The employed
 values for dimensionless entropy (terminal Lorentz factor)
 and kinetic luminosity are chosen
 as $\eta_0=400$ and  $L_0=10^{53}~{\rm erg/s}$ for the spine.
 The purple and black lines show the cases of the  dimensionless entropy of the sheath
 given by   $\eta_1=200$  and  $\eta_1=100$, respectively.   
 In each case, the kinetic luminosity of the sheath is
 given by $L_1=(\eta_0/\eta_1)L_0$.
}  
\label{nuave}
\end{center}
\end{figure}

%In Fig. \ref{CRave},
%we show the results of the  analysis for the photons observed 
%by the observer located at $\theta_{\rm obs}=\theta_0=0.5^{\circ}$.
% 
%For an illustrative purpose,
In Fig. \ref{CRave}, we show the distribution of
the average number of crossings for a given observed photon energy,
$\langle n_{\rm cr} \rangle_{\nu}$.
On the other hand, distribution of the average observed energy
for a given number of crossings, $\langle \nu \rangle_{n_{\rm cr}}$,
is displayed in Fig. \ref{nuave}.
%which is calculated by 
%adding up the number of event for all the photons at the energy 
%and then dividing it by the number of the corresponding photons.
%is simply calculated by 
%adding up the number of event for all the photons at the energy 
%and then deviding it by the number of the corresponding photons.
%
%On the other hand, we plot the distribution of 
%the average energy of photons for a given number of crossings in
%Fig. \ref{nuave}, 
%which is calculated by sum of the energy of photons having 
%
%In Figs. \ref{CRave} and \ref{nuave}, we show the distribution of 
%the average number of crossings for a given photon energy and
%the  the average energy of photons for a given number of crossings,
% respectively.
The two cases of stratified jet that are
displayed in the figures by the purple
(Case I: $\eta_0=400$ and $\eta_1 = 200$) and black lines
 (Case II: $\eta_0=400$ and $\eta_1 = 100$) correspond to the 
analysis of photons  shown in Fig. \ref{G400compare} using same colors.

\begin{figure*}[htbp]
\begin{tabular}{cc}
\hspace{-0.5cm}
\begin{minipage}{0.5\hsize}
\begin{center}
\includegraphics[width=9cm,keepaspectratio,clip]{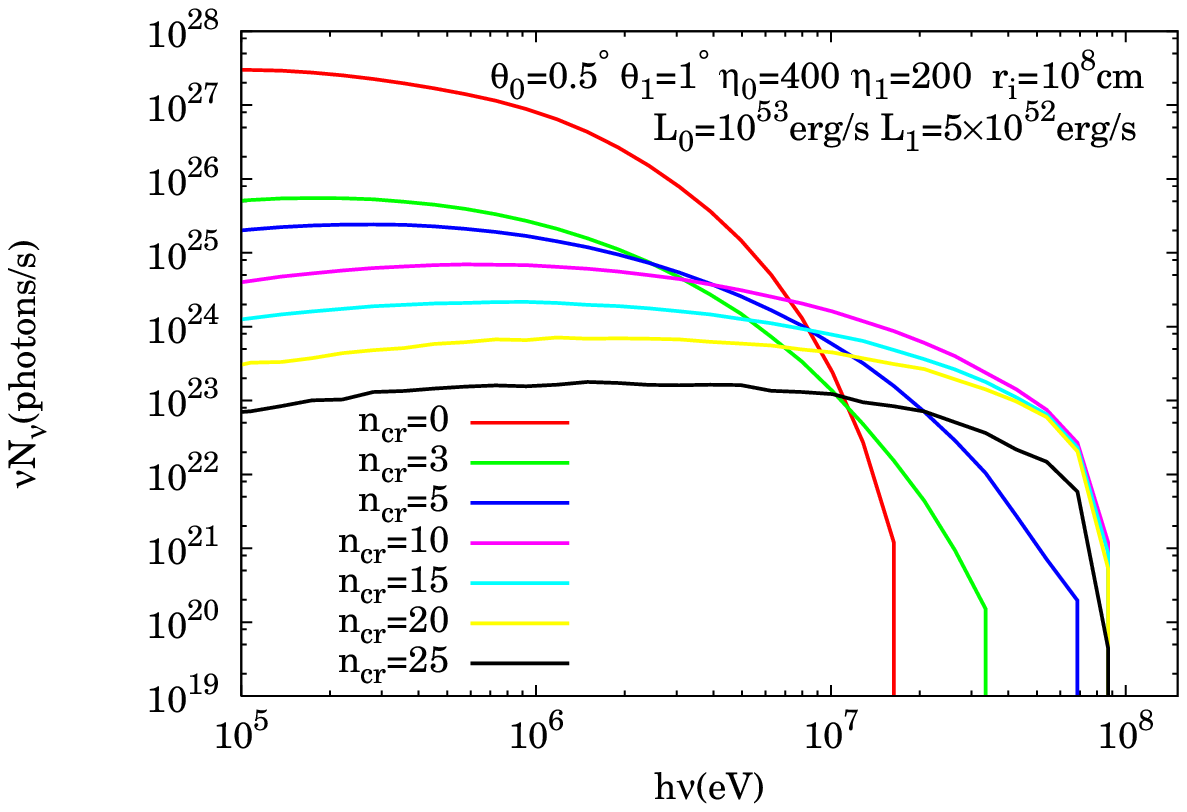}
\end{center}
\end{minipage}
\begin{minipage}{0.5\hsize}
\begin{center}
\includegraphics[width=9cm,keepaspectratio,clip]{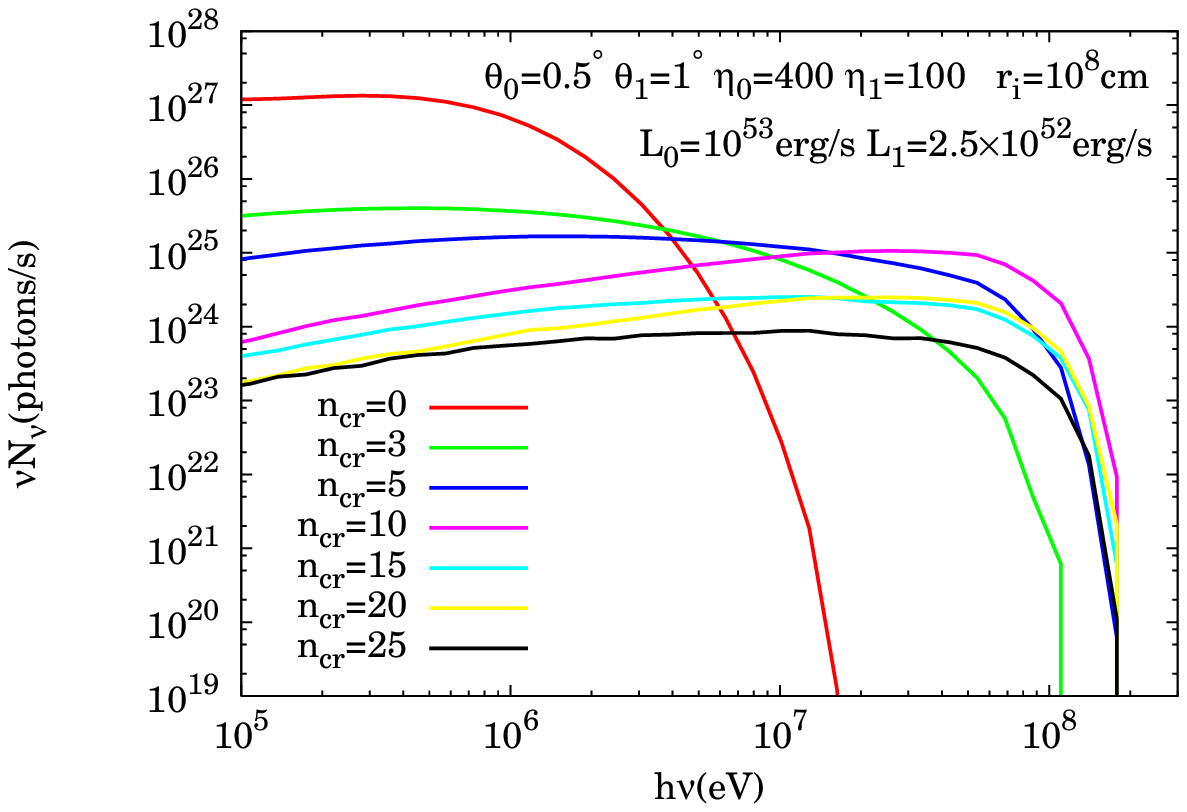}
\end{center}
\end{minipage}
\end{tabular}
\caption{
Energy distribution of photons
for a given number of crossing events
 in the case of jet
 in which a spine region with half opening angle of $\theta_0=0.5^{\circ}$
 is embedded in a wider sheath outflow with half opening angle of
 $\theta_1=1^{\circ}$.
 The red, green, blue, purple, light blue, yellow and black lines
 display the photons 
 that have experienced $n_{\rm cr} = 0$, $3$, $5$, $10$, $15$, $20$ and
 $25$ crossings, respectively.
 The employed
 values for dimensionless entropy (terminal Lorentz factor)
 and kinetic luminosity are chosen
 as $\eta_0=400$ and  $L_0=10^{53}~{\rm erg/s}$ for the spine.
 The left and right panels show the cases
 where the dimensionless entropy of the sheath is
 given by   $\eta_1=200$  and  $\eta_1=100$, respectively.   
 In each case, the kinetic luminosity of the sheath is
 given by $L_1=(\eta_0/\eta_1)L_0$.
}
\label{Ndis}
\end{figure*}

From  Fig. \ref{CRave}, it is confirmed that the photons at
higher energies tend to have larger number of crossings. % can be confirmed.
While the photons below the thermal peak energy
($h\nu \lesssim ~{\rm MeV}$) do not require multiple crossings, 
 the prominent non-thermal component
 extending above $\sim 1~{\rm MeV}$ is produced by
the photons that
cross the boundary layer $\sim 10-15$ times in average.
Comparing the two cases, the overall distribution of 
$\langle n_{\rm cr} \rangle_{\nu}$.
does not vary much.
The difference in the average number of crossing is within $\sim 2$ 
in all energies up to $\sim 100~{\rm MeV}$.
Note that, however, that this does not imply that
the average energy for a given number of crossings
 $\langle \nu \rangle_{n_{\rm cr}}$ does not vary much in the two cases.
Conversely, the
difference in  $\langle \nu \rangle_{n_{\rm cr}}$
is quite large between the two cases as is seen in Fig. \ref{nuave},
 since the average energy gain per crossing
is quite sensitive to the 
ratio in the terminal Lorentz factor $\Gamma_0/\Gamma_1$
(see Eqs. (\ref{up}) and (\ref{down})).
However,
due to the large dispersion in the  energy gain per crossing, 
%show large dispersion around the average value,
%due to the large dispersion in the energy gain per crossings,
%since the dispersion of the energy for a given number of scattering is large, 
the energy distribution of photons does not show a sharp peak
at the average energy $\langle \nu \rangle_{n_{\rm cr}}$
but extends to energies below and above $\langle \nu \rangle_{n_{\rm cr}}$
by many orders of magnitude.
As a result,
the distribution of $\langle n_{\rm cr} \rangle_{\nu}$
do not directly 
reflect the distribution of  $\langle \nu \rangle_{n_{\rm cr}}$,
since the 
 photons which have crossed the boundary layer a certain number of times
 can dominate over the other population of photons 
in wide energy ranges.
To clarify this, we show the
 energy distribution of the photon number count rate, $\nu N_{\nu}$
[photons/s], for a given number of crossings in \ref{Ndis}.
From the figure,
it is confirmed that, while there is 
a significant discrepancy in $\langle \nu \rangle_{n_{\rm cr}}$,
the photons with number of crossing $n_{\rm cr}\sim 10$ tend
to dominate the population in the energy range
in which non-thermal component becomes prominent
($h \nu \gtrsim {\rm few}~{\rm MeV}$) in both cases.
For this reason, the resultant distribution of
$\langle n_{\rm cr} \rangle_{n_{\rm cr}}$ does not vary much in the two cases.

Regarding the distribution of the average energy,
 $\langle \nu \rangle_{n_{\rm cr}}$ tends to increase %monotonically 
with the increasing number of crossing $n_{\rm cr}$ initially
and then approaches a constant value.
This asymptotic behaviour is
due to the Klein-Nishina effect.
As the photon energy becomes large and exceeds
$100~{keV}$ in the comoving frame, acceleration efficiency is reduced
by the effect (see \S\ref{stratified} for detail), and 
the average energy can no longer increase.
%\bl{
As in seen in Fig. \ref{nuave},
the dependence of $\langle \nu \rangle_{n_{\rm cr}}$ 
on $n_{\rm cr}$ is not smooth but rather  bumpy.
This reflects the 
fact that the photons tend  to be upscattered when 
crossing from the sheath to spine region occurs (see Eq. (\ref{up})),
 while, on the other hand,
tend to be downscattered when crossing in the opposite direction occurs (see Eq. (\ref{down})).
Therefore, the energy of individual photons
 is not a monotonically increasing function of $n_{\rm cr}$ but 
shows a bumpy dependence,
since upscattering and downscattering occurs alternately in 
each crossing events.
Although somewhat reduced,
this feature remains even after averaging up and 
leads to the appearance wiggles in the
distribution of  $\langle \nu \rangle_{n_{\rm cr}}$.
% 
%}

\begin{figure*}[htbp]
\begin{tabular}{cc}
\hspace{-0.5cm}
\begin{minipage}{0.5\hsize}
\begin{center}
\includegraphics[width=9cm,keepaspectratio,clip]{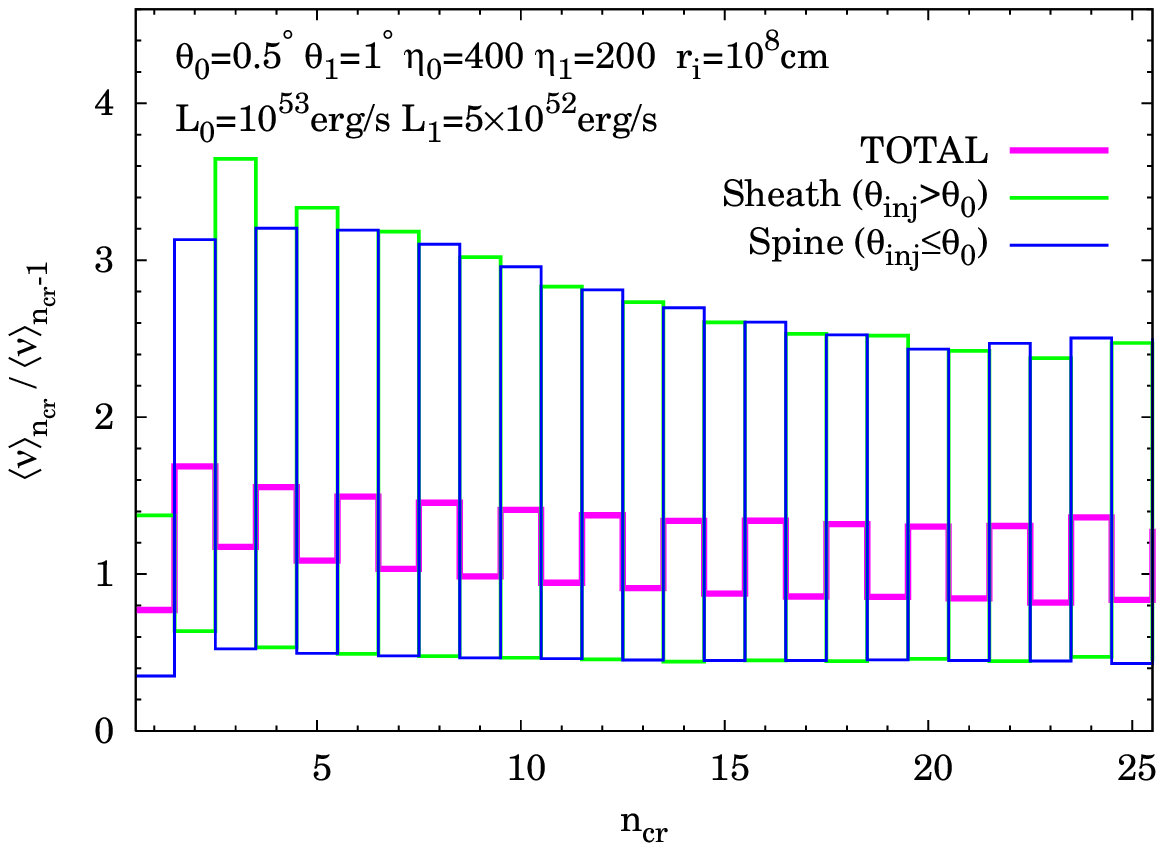}
\end{center}
\end{minipage}
\begin{minipage}{0.5\hsize}
\begin{center}
\includegraphics[width=9cm,keepaspectratio,clip]{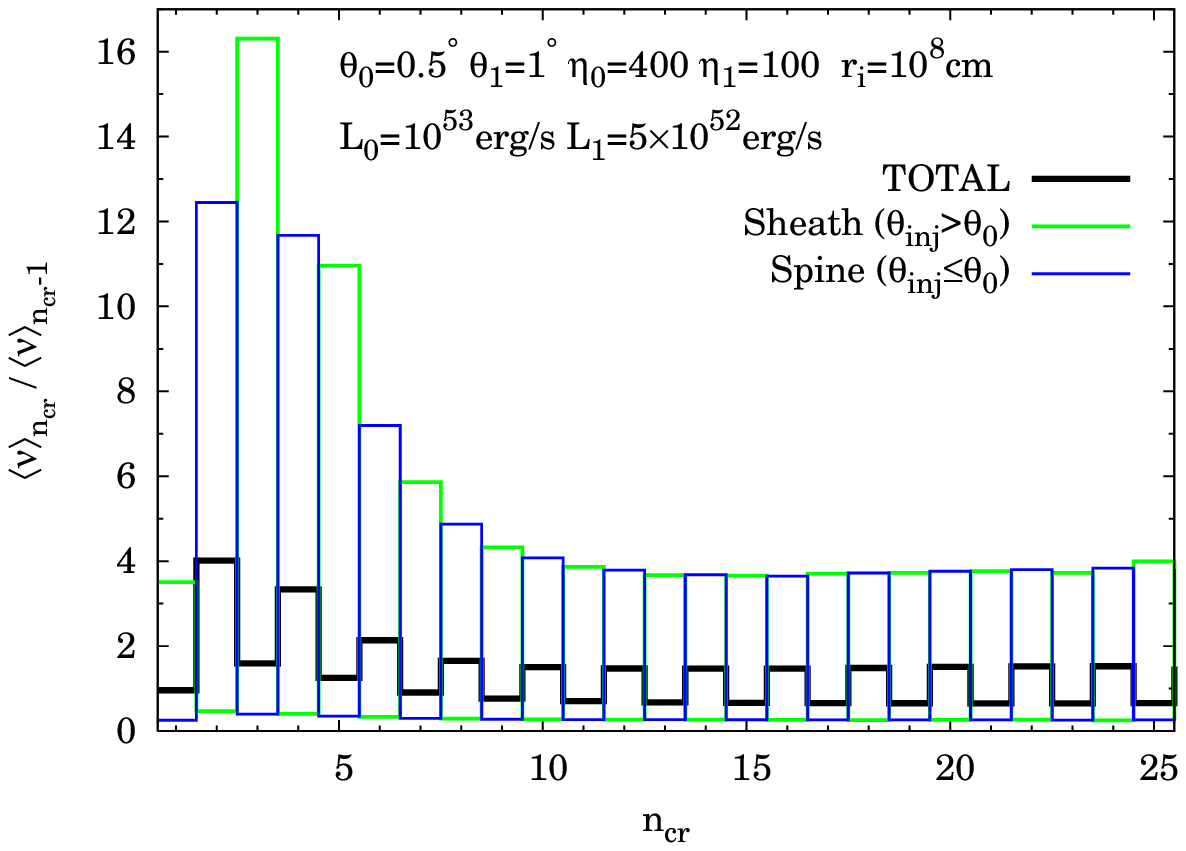}
\end{center}
\end{minipage}
\end{tabular}
\caption{
Ratio of the average observed energy for 
photons which cross the spine-sheath boundary layer 
 ($\theta=\theta_0$)
 $n_{\rm cr}$ times to those which experience
 $n_{\rm cr}-1$ crossings for a jet
 in which a spine region with half opening angle of $\theta_0=0.5^{\circ}$
 is embedded in a wider sheath outflow with half opening angle of
 $\theta_1=1^{\circ}$.
 The employed
 values for dimensionless entropy (terminal Lorentz factor)
 and kinetic luminosity are chosen
 as $\eta_0=400$ and  $L_0=10^{53}~{\rm erg/s}$ for the spine.
 The left and right panels show the cases of the  dimensionless entropy of the sheath
 given by   $\eta_1=200$  and  $\eta_1=100$, respectively.   
 In each case, the kinetic luminosity of the sheath is
 given by $L_1=(\eta_0/\eta_1)L_0$.
 The energy ratio for total photons are displayed 
 in the left and right panels with
 the purple and black lines, respectively.
 In addition, also the energy ratio for the 
 photons that were initially injected in the spine region
 ($\theta_{\rm inj}\leq \theta_0$)  and 
 sheath region  ($\theta_{\rm inj}>\theta_0$) are  
 shown by blue and green lines, respectively, in both panels.
}
\label{nurat}
\end{figure*}

To quantify the energy gain and loss
 in each crossing,
we display the ratio of the average energy 
for photons with $n_{\rm cr}$ crossings 
to that for photons with  $n_{\rm cr}-1$ crossings
($\langle \nu \rangle_{n_{\rm cr}}/ \langle \nu \rangle_{n_{\rm cr}-1}$)
in Fig. \ref{nurat}.
In addition to the analysis of the total photons,
we also display the results of the analysis
 for the two populations of
 photons that were initially injected 
in the spine region
($\theta_{\rm inj}\leq \theta_0$; {\it blue line})
 and sheath regions ($\theta_{\rm inj}> \theta_0$; {\it green line}).
In the former case ($\theta_{\rm inj}\leq \theta_0$),
if $n_{\rm cr}$ is an odd (even) number,
the number of crossings from the spine (sheath) to sheath (spine) region
is greater by $1$ than that for $n_{\rm cr} - 1$ crossings.
%the number of crossings from the spine (sheath) to sheath (spine) region
%for the photons with odd (even) number of $n_{\rm cr}$ 
%is greater by $1$
%than those with  $n_{\rm cr} - 1$.
%
Hence, the number of downscattering (upscattering) event 
is greater by $1$
for the photons with an odd (even) number of $n_{\rm cr}$ 
than for those with  $n_{\rm cr} - 1$ crossings.
As a result,
 $\langle \nu \rangle_{n_{\rm cr}}/ \langle \nu \rangle_{n_{\rm cr}-1}$ 
is less (greater)
than unity for the odd (even) number of $n_{\rm cr}$. 
On the other hand, as is obvious,
the opposite is true for the latter case ($\theta_{\rm inj}> \theta_0$).
Jets with a larger difference in the terminal Lorentz factor
(Case II; {\it right panel}) show
a larger range of energy ratio
$\langle \nu \rangle_{n_{\rm cr}}/ \langle \nu \rangle_{n_{\rm cr}-1}$
than those with a smaller difference 
(Case I; {\it left panel}), 
since the efficiency of the single upscattering (downscattering)
increases (decreases) as the relative difference in the Lorentz factor 
becomes larger.

Regarding the upscatterings,
the energy ratio 
$\langle \nu \rangle_{n_{\rm cr}}/ \langle \nu \rangle_{n_{\rm cr}-1}$
for  photons that cross the boundary from the sheath to spine region
 (green line)
only once ($n_{\rm cr}=1$) is relatively small
 because a
large fraction of these photons experience the crossing when 
the velocity shear is not fully developed ($r<r_{\rm s0}$).
For  photons with a larger number of crossings ($n_{\rm cr}\gtrsim 2$),
%the radius at which the last crossing event takes place increases, and 
a large fraction of the photons experience the last crossing
at $r>r_{\rm s0}$, where the velocity shear is fully developed.
Therefore, the energy ratio is larger than that for $n_{\rm cr}=1$ and 
is roughly constant as long as the  Klein-Nishina effect is negligible.
When the  Klein-Nishina effect becomes important
($n_{\rm cr} \gtrsim 10$ for Case I and $n_{\rm cr} \gtrsim 5$ for Case II),
$\langle \nu \rangle_{n_{\rm cr}}/ \langle \nu \rangle_{n_{\rm cr}-1}$
decreases as $n_{\rm cr}$ increases, and again asymptotically approaches a constant value. 
On the other hand, 
regarding the case of downscatterings,
the energy ratio is relatively insensitive to
the number of crossings in both cases
 and is roughly in the range $\sim 0.25-0.6$.
It is worth noting that the values of energy ratio 
are  consistent  within a factor of $\sim 2$ 
with the rough estimations given by Eqs. (\ref{up}) and (\ref{down}).
Above the saturation radius ($r \geq r_{\rm s0}$),
the equations predict an upscattering (downscattering) with energy ratio of
$\sim 2.5$ ($\sim 0.6$) and $\sim 8.5$ ($\sim 0.5$) for Cases I and II, 
respectively.
The energy ratio of the total photons
is larger and smaller when the photons from the spine correspond to 
the upscattering (even $n_{\rm cr}$) and downscattering (even $n_{\rm cr}$),
respectively,
since more photons  originate from the spine region 
than from the sheath region.

\begin{figure*}[htbp]
\begin{tabular}{cc}
\hspace{-0.5cm}
\begin{minipage}{0.5\hsize}
\begin{center}
\includegraphics[width=9cm,keepaspectratio,clip]{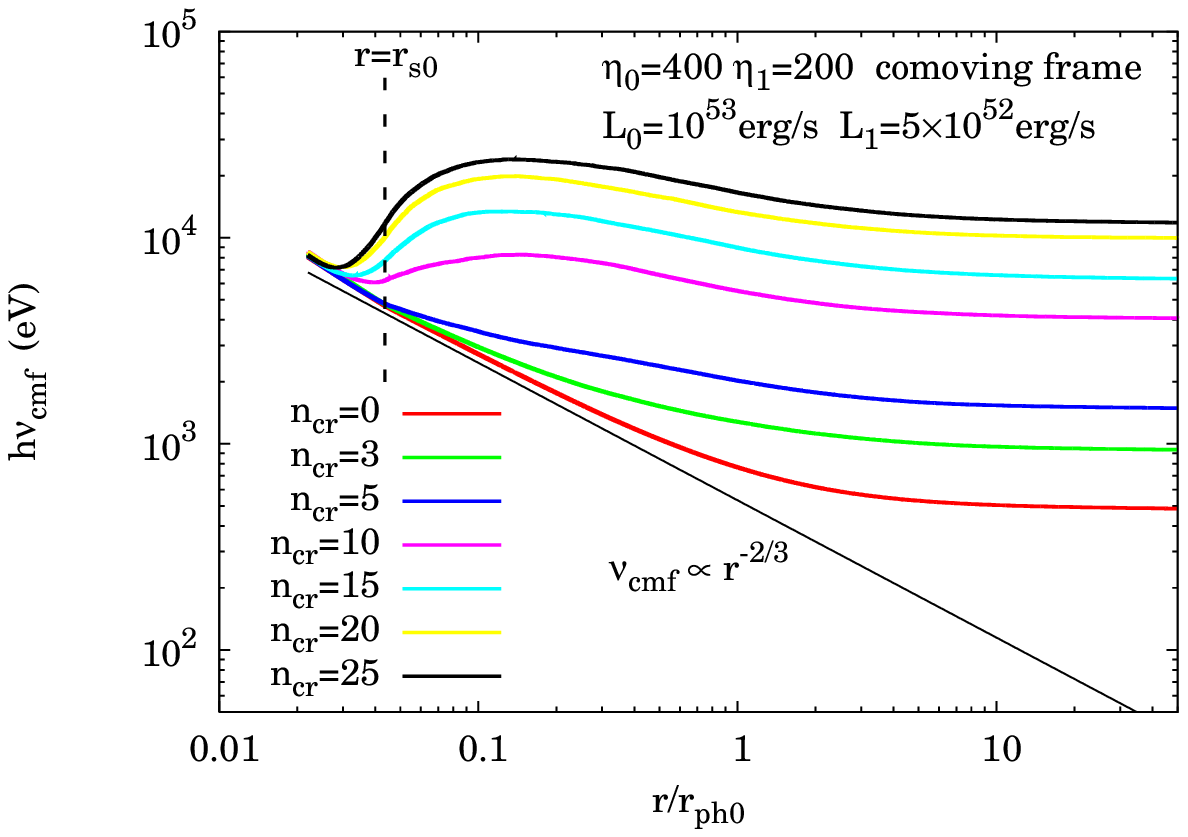}
\end{center}
\end{minipage}
\begin{minipage}{0.5\hsize}
\begin{center}
\includegraphics[width=9cm,keepaspectratio,clip]{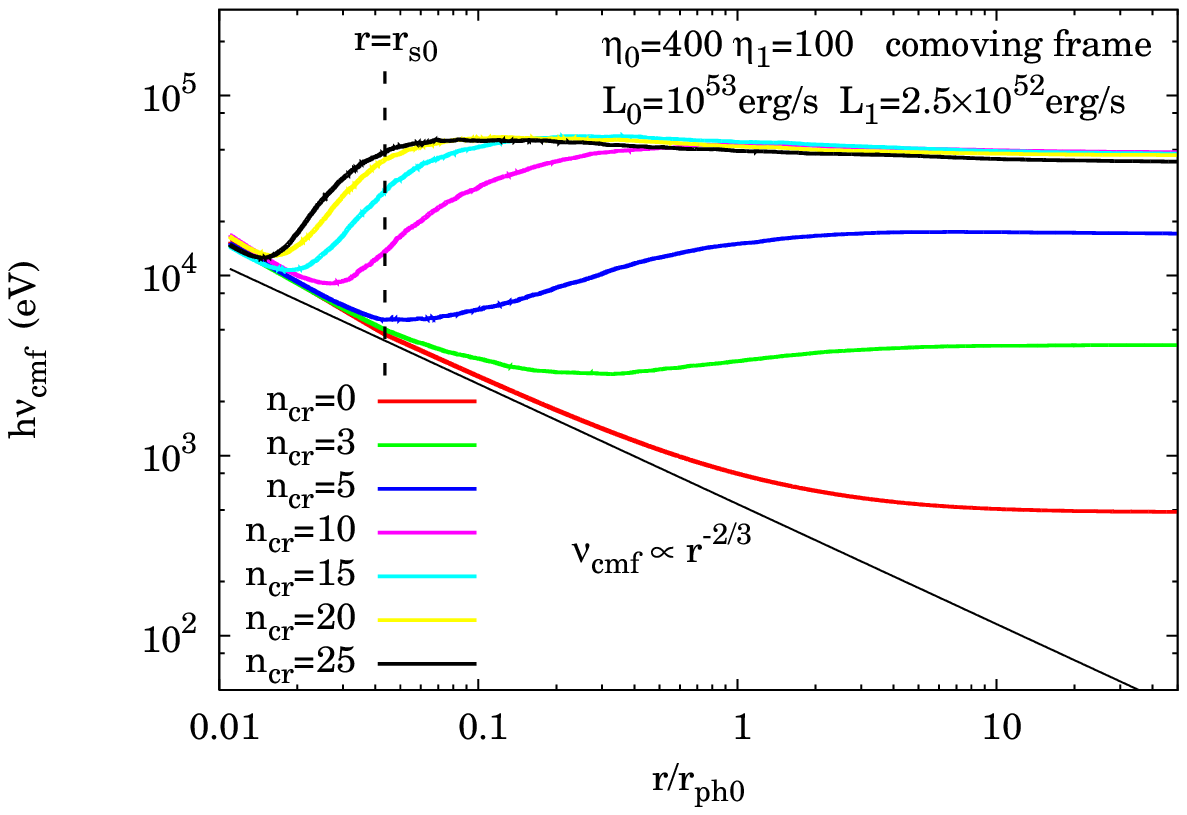}
\end{center}
\end{minipage}
\end{tabular}
\caption{
Average comoving energy of photons as a function of radius
 for a jet
 in which a spine region with half opening angle of $\theta_0=0.5^{\circ}$
 is embedded in a wider sheath outflow with half opening angle of
 $\theta_1=1^{\circ}$.
 The red, green, blue, purple, light blue, yellow and black lines
 display the photons 
 that have experienced $n_{\rm cr} = 0$, $3$, $5$, $10$, $15$, $20$ and
 $25$ crossings, respectively.
 The thin black solid line plots the curve of $\propto r^{-2/3}$,
 which corresponds to the adiabatic cooling expected 
 above the saturation radius ($r\geq r_{\rm s0}$).
 The employed
 values for dimensionless entropy (terminal Lorentz factor)
 and kinetic luminosity are chosen
 as $\eta_0=400$ and  $L_0=10^{53}~{\rm erg/s}$ for the spine.
 The left and right panels show the cases of the dimensionless entropy of the sheath
 given by   $\eta_1=200$  and  $\eta_1=100$, respectively.   
 In each case, the kinetic luminosity of the sheath is
 given by $L_1=(\eta_0/\eta_1)L_0$.
}
\label{nucmfevo}
\end{figure*}

Lastly,
to obtain a further insight
into the relation between the photon acceleration and number of crossing,
we show the 
evolution of the
average photon energy 
evaluated in the comoving frame, $\nu_{\rm cmf}$,
 with radius  in Fig \ref{nucmfevo}.
 The red, green, blue, purple, light blue, yellow and black lines
 display the photons 
 that have experienced $n_{\rm cr} = 0$, $3$, $5$, $10$, $15$, $20$ and
 $25$ crossings, respectively.
 For comparison, we also 
 plot the curve of $\nu_{\rm cmf} \propto r^{-2/3}$
 with a thin black line,
 which corresponds to the adiabatic cooling expected 
 above the saturation radius ($r\geq r_{\rm s0}$).

 Regarding the photons that do not experience any crossings ($n_{\rm cr}=0$),
 the overall evolution of the energy is
 determined solely by the adiabatic cooling due 
 to the expansion of the jet, since photon acceleration does not take place.
 Below the saturation radius of the spine $r \leq r_{\rm s0}$,
 the cooling rate in the spine region ($\nu_{\rm cmf} \propto r^{-1}$)
 is higher than that in the
 sheath region ($\nu_{\rm cmf} \propto r^{-2/3}$), since
 the spine region is in the acceleration phase, while 
 the sheath region is in the coasting phase ($r\geq r_{\rm s1}$).
% As a result, the
 Hence, the energy evolution of 
 the average comoving  energy 
 is in between the two cooling rates.
 At larger radii ($r \geq r_{\rm s0}$),
 since  both regions are in the coasting phase,
 the average energy evolves as
 $\nu_{\rm cmf} \propto r^{-2/3}$ until
 the coupling between the photon and matter becomes weak
 ($r\lesssim 0.1 r_{\rm ph0}$).
 The cooling rate gradually reduces at $r\gtrsim 0.1 r_{\rm ph0}$
 due to the weak coupling, and the comoving energy
 approaches a constant value.
 It is worth noting that 
 the behaviour of the average comoving energy for photons %($n_{\rm cr}=0$) 
 above the  saturation radius is consistent
 with that found in the previous studies \citep[e.g.,][]{P08,  B11, BSV13}.

 Regarding the photons with at least one crossing ($n_{\rm cr}\geq 1$),
 the evolution of the comoving energy cannot be described 
 only by the adiabatic cooling due to the presence of the photon acceleration.
 As  the number of the crossings increases, the departure 
 from the simple adiabatic cooling becomes significant
 and the comoving energies tend to be larger.
 For a given number of crossings, the departure is more prominent
 in Case II than in Case I
 due to the increase in the acceleration efficiencies.
 When the number of the crossings is sufficiently large so that the effect 
 of Klein-Nishina becomes non-negligible, the evolution of the 
 average comoving energy asymptotically approaches a single curve since
 the acceleration saturates.
 This tendency is clearly seen  in Fig. \ref{nucmfevo}
 (for example, see lines  that display the 
 evolution of photons with $n_{\rm cr} \geq 10$ in Case II).

%\begin{figure}[ht]
%\begin{center} 
%\includegraphics[width=9cm]{f8.eps}
%\caption 
%{Spe 
%}  
%\label{G4-2csp}
%\end{center}
%\end{figure}

%\begin{figure}[ht]
%\begin{center} 
%\includegraphics[width=9cm]{f9.eps}
%\caption 
%{Spe 
%}  
%\label{G4-1csp}
%\end{center}
%\end{figure}

%\begin{figure}[ht]
%\begin{center} 
%\includegraphics[width=9cm]{f10.eps}
%\caption 
%{Spe 
%}  
%\label{G4-2nurat}
%\end{center}
%\end{figure}

%\begin{figure}[ht]
%\begin{center} 
%\includegraphics[width=9cm]{f11.eps}
%\caption 
%{Spe 
%}  
%\label{G4-1nurat}
%\end{center}
%\end{figure}

%\begin{figure}[ht]
%\begin{center} 
%\includegraphics[width=9cm]{f12.eps}
%\caption 
%{Evolution of 
%}  
%\label{G4-2evo}
%\end{center}
%\end{figure}

%\begin{figure}[ht]
%\begin{center} 
%\includegraphics[width=9cm]{f13.eps}
%\caption 
%{Evolution of 
%}  
%\label{G4-1evo}
%\end{center}
%\end{figure}

%\newpage

%\twocolmn

\begin{figure}[ht]
\begin{center} 
\includegraphics[width=9cm]{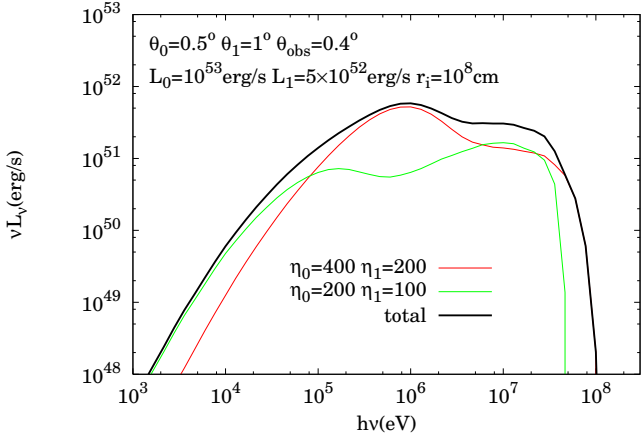}
\caption 
{Observed luminosity spectrum in the case of spine-sheath jet
 in which a spine jet with half opening angle of $\theta_0=0.5^{\circ}$
 is embedded in a wider sheath outflow with half opening angle of
 $\theta_1=1^{\circ}$.
 The observer angle is fixed as $\theta_{\rm obs} = 0.4^{\circ}$.
 The black solid line shows the overall spectrum
 for 
 the case when the dimensionless entropy of the 
 spine-sheath jet have evolved from $\eta_0 = 400$ and $\eta_1=200$ (initial stage)
 to  $\eta_0 = 200$ and $\eta_1=100$ (later stage), while
 the kinetic luminosity  is
 fixed as $L_0 = 10^{53}~{\rm erg/s}$ and $L_1= 5\times 10^{52}~{\rm erg/s}$.
 The red and green solid lines show the 
 contribution from the initial and later stages, respectively. 
} 
\label{G400spose}
\end{center}
\end{figure}

%\begin{figure}[ht]
%\begin{center} 
%\includegraphics[width=9cm]{f7.eps}
%\caption 
%{Observed luminosity spectrum in the case of spine-sheath jet
% in which the spine jet with half opening angle of $\theta_0=0.5^{\circ}$
% is embedded in a wider sheath outflow with half opening angle of
% $\theta_1=1^{\circ}$.
%}  
%\label{G200spose}
%\end{center}
%\end{figure}

\begin{figure*}[ht]
\begin{center} 
\includegraphics[width=12cm]{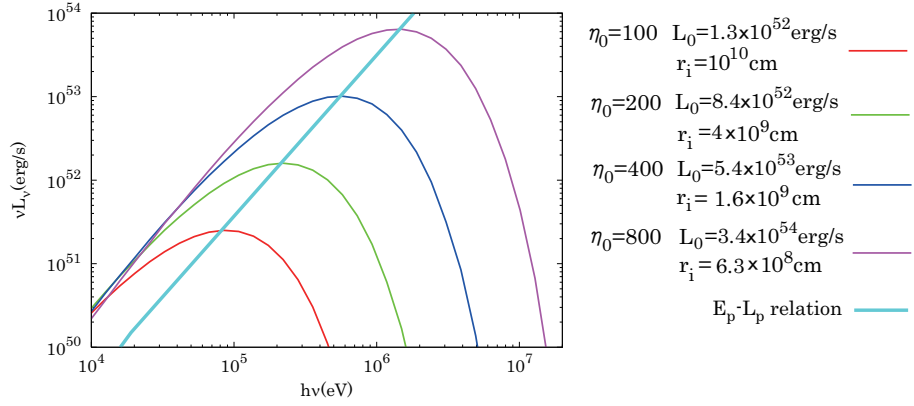}
\caption 
{Observed  spectra in the case of a uniform jet that reproduces
  the observed $E_{\rm p}-L_{\rm p}$ relation and radiative efficiency of
 $\eta_{\gamma}\sim 0.2$.
The red, green, blue, purple lines correspond to the cases of
$\eta_0 = 100$, $\eta_0 = 200$, $\eta_0 = 400$ and $\eta_0 = 500$.
The thick light blue line displays the observed $E_{\rm p}$-$L_{\rm p}$ relation.
}  
\label{EpLpfig}
\end{center}
\end{figure*}

\section{DISCUSSIONS}
\label{discussions}
%In this section we discuss our result and its implications. 

%In the previous section, we have demonstrated
%that the  Band function can be reproduced when a stratified 
%jet with strong velocity shear in the boundary layer is considered.
%Here we discuss the validity and the implication of our results.

\begin{figure}[ht]
\begin{center} 
\includegraphics[width=9cm]{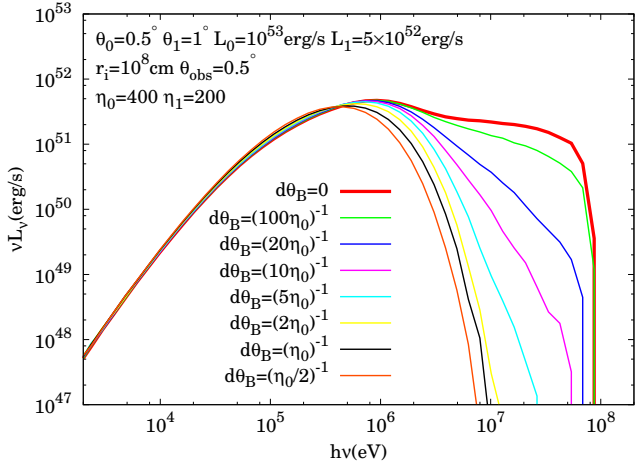}
\caption 
{
Observed luminosity spectrum
for a spine-sheath jet in which the width of the boundary layer
 $d\theta_{\rm B}$ is taken to be non-zero.
The the spine and sheath regions are 
located in the transverse range
$0\leq \theta \leq \theta_{\rm i}=\theta_0 - d\theta_{\rm B}/2$ and 
$\theta_{\rm ii}=\theta_0 + d\theta_{\rm B}/2 \leq \theta \leq \theta_1$,
respectively.
In all cases,
fixed values of $\theta_0 = 0.5^{\circ}$ and  $\theta_1 = 1^{\circ}$
are employed and  the observer angle 
is fixed at the midpoint of the boundary layer
 $\theta_{\rm obs}=\theta_0 = 0.5^{\circ}$.
The employed
values for dimensionless entropy (terminal Lorentz factor)
and kinetic luminosity are chosen
as $\eta_0=400$ and  $L_0=10^{53}~{\rm erg/s}$ for the spine and
 $\eta_0=200$ and  $L_0=5\times 10^{52}~{\rm erg/s}$ for the sheath.
While the thick red line displays the case for an infinitesimal width
 ($d\theta_{\rm B}=0$),
the thin green, blue, purple, light blue, yellow, black and red lines 
display the cases for widths of
 $d\theta_{\rm B}=(100\eta_0)^{-1}$,  
 $(20\eta_0)^{-1}$,  
 $(10\eta_0)^{-1}$,  
 $(5\eta_0)^{-1}$,  
 $(2\eta_0)^{-1}$,  
 $\eta_0^{-1}$
and  $(\eta_0/2)^{-1}$, respectively.
}  
\label{G42sm}
\end{center}
\end{figure}

\begin{figure}[ht]
\begin{center} 
\includegraphics[width=9cm]{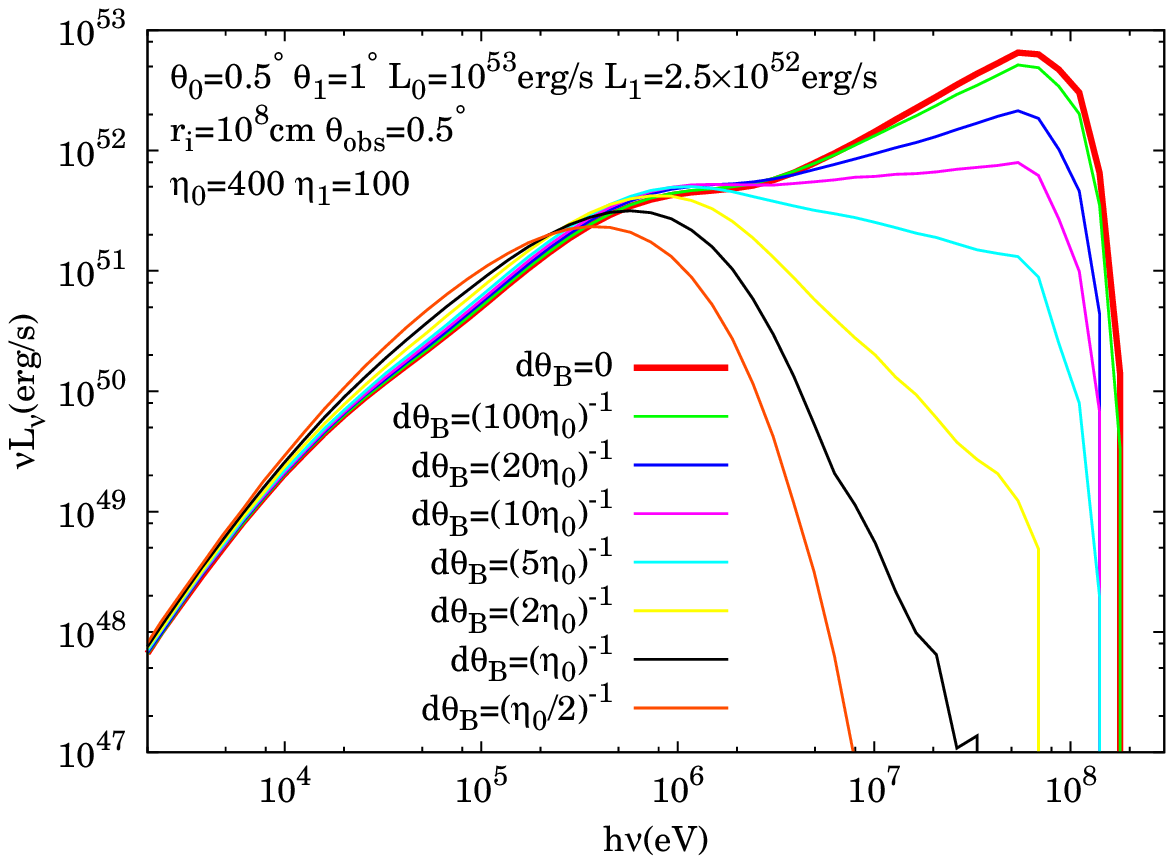}
\caption 
{
Same as Fig.\ref{G42sm}, but for
  $\eta_1=100$,
  $L_1=2.5 \times 10^{52}~{\rm erg/s}$.
}  
\label{G41sm}
\end{center}
\end{figure}

\subsection{Comparison with Observations}

 So far, we have shown that in the presence 
 of velocity shear,
 photons can be accelerated at the shear region
 producing a high energy power-law component above the thermal peak energy.
 From observations, the observed spectra of GRB prompt emission can be
 often modeled by
 a Band function,
  which is a smoothly joined broken power-law that peaks at $\sim$ a few
 $100~{\rm keV}$.
 The photon indices below and above the peak ($\alpha_{\rm ph}$ and $\beta_{\rm ph}$) vary
 from source to source
 ($\nu L_{\nu}\propto \nu^{\alpha_{\rm ph} + 2}$ for $\nu < \nu_{\rm p}$
  and $\nu L_{\nu}\propto \nu^{\beta_{\rm ph} + 2}$ for $\nu > \nu_{\rm p}$).
 Focusing on the high-energy slope $\beta_{\rm ph}$,
 the measured values are roughly in the range between $-2$ and $-3$,
 with a typical value at  $\sim - 2.5$.  
%
% As shown above, 
 The power-law component predicted in our model can
 accommodate various spectral indices  depending 
 on the flow parameters as well as observer angle. 
%
% The high energy slope of the observed GRBs also vary significantly 
% from source to source
%
 Hence,   by adopting appropriate values for our parameter set,
 the
 observed high energy spectral slope
 ($\nu L_{\nu} \propto \nu^{-0.5 \pm 0.5} $)
 as well as the harder and softer slopes observed in some GRBs 
 can be reproduced 
 as shown in Figs. \ref{G400st}-\ref{G400compare}.
%
% It is also noted that our model can also accomodate harder as well as
% softer slopes observed in some GRBs.
%!

 Interestingly,
 our model can also reproduce spectral features seen in
 some peculiar GRBs such as
 GRB 090510\citep{AAA10}, 090902B\citep{AAA09} and 090926A\citep{AAA11}. 
 In these bursts, in addition to a Band-type (or thermal-like) component,
 an extra hard power-law component (photon index larger than $-2$)
 is required to model the overall spectrum. 
 As shown in Fig. \ref{G400compare},
 in some cases,
 especially when the difference between  $\eta_0$ and $\eta_1$ is large, 
 the non-thermal power-law component
 does not join the thermal component smoothly at the peak energy,
 but appears as a 
 hard power-law component which
 becomes prominent at energies above the bump  of the thermal component.
 These spectral features resemble those found in these bursts.
 Hence, we emphasize that the spectra of peculiar bursts may be
 due to the
 relatively large difference in $\eta$ (or bulk Lorentz factor)
 in the shear region.
 It is noted, however, that the
 extra power-law component that also
 extends to energies %not only above but also
 below the
 thermal-like component found in GRB 090902B and 090510   
 is hard to explain.
 If present, 
 different emission components 
 such as synchrotron emission as discussed in \citet{PZR12}  may be required
 in order to explain the low energy end of the power-law component.
 We also note that, in order to extend the power-law component up to
 ${\rm GeV}$ energies, quite large Lorentz factors are required,
 since the non-thermal component extends only up to energy of 
 $h \nu \lesssim 200(\eta / 400)~{\rm MeV}$ due to
 the Klein-Nishina effect.
 A detailed discussion on this issue is given in 
 \S\ref{GeV}.

 On the other hand, the spectral slope below
 the peak energy
 is only moderately sensitive to 
 the values of the parameters
 and can be well approximated by a blackbody emission
 from the off-axis region ($|\theta - \theta_{\rm obs}|\gtrsim \Gamma^{-1}$)
 of a sphere at a radius $r\sim r_{\rm ph}/5$ as in the case of 
 a uniform jet.
% which 
% is given by Eq. (\ref{photoana}).
 In all cases, the low energy photon indices are 
 roughly  $\alpha_{\rm ph} \sim 0.5$  which
 is harder than the typical observed value ($\alpha_{\rm ph} \sim -1$).
 It is noted, however, that
 this does not imply that 
 the low energy slope cannot be reproduced within this scenario.
 For instance,
 while the instantaneous spectrum is hard,
 the time integration can lead to a softer spectrum
 if the evolution history of the outflow is considered.
 To demonstrate this, let us consider
 the case when the dimensionless entropies of the 
 spine-sheath jet have evolved from $\eta_0 = 400$ and $\eta_1=200$
 (initial stage)
 to  $\eta_0 = 200$ and $\eta_1=100$
 (later stage), while
 the kinetic luminosity  is
 fixed as $L_0 = 10^{53}~{\rm erg/s}$ and $L_1= 5\times 10^{52}~{\rm erg/s}$.
 We illustrate the resultant
 spectrum in Fig. \ref{G400spose} 
 under the assumption that
 a nearly equal time period
 is spent in the two stages.
 The black solid line shows the overall spectrum, 
 and the red and green solid lines show the 
 contribution from the initial and 
 later stages, respectively. 
 Since the peak energy and the luminosity 
 vary  with the dimensionless entropy of the flow
 roughly as $\propto \eta^{8/3}$,
 they are smaller by a factor $\sim 6$,
 in the later stage.
 As a result, 
 while the peak energy and luminosity of the
 overall spectra is determined by the initial stage,
 at lower energies, contribution from the 
 later stage becomes dominant.
 Due to the superposition of the two components,
 the spectrum
 below the peak becomes softer than
 that of the individual component, so that a
 typical observed spectrum  ($\nu L_{\nu} \propto \nu$)
 is reproduced.
Note also that the high energy part
 of the overall 
 spectrum  largely resembles typical ones from observations.
 Hence, we conclude that typical observed spectra can 
 be successfully reproduced when 
 a  time evolution is considered.
\footnote{
 The  multi-temperature effect due to 
 the continuous change of the flow properties in the
 $\theta$ direction may
 also be a possible origin for the
 soft low energy slope \citep{LPR13}.}  

% As an alternative scenario, 
% the  multi-temperature effect due to 
% the continuous change of the flow properties in
% lateral direction may
% also be a possible origin for the 
% soft low energy slope.
% This 
% was shown in the  study by \citet{LPR12}
% which explored the detail of photospheric emission
%% from a outflow having smooth change
% of fluid properties in lateral direction. 
%
% If this is the case, 
% total spectra of typical GRBs may be explained by 
% considering a smooth flow profile in addition
% to the discontinous change.

\subsection{On the Observer Angle Dependence and Structure of the Jet}
As shown in the previous section, our results have a quite
strong dependence on the observer angle $\theta_{\rm obs}$.
While strong non-thermal emission can be seen 
when the observer angle is close to the 
angle in which a velocity shear is present
 ($|\theta_0 -\theta_{\rm obs}| \lesssim \Gamma^{-1}$),
the non-thermal feature tends to be weaker
for observer angles far from $\theta_0$.
This tendency seems to contradict the fact that 
most of the observed GRBs have non-thermal features in 
their spectra.
%Therefore, it is obvious that
%simple steady state jet with spine-sheath structure
%cannot explain prompt emissions.
However, we emphasize that 
%this does not imply that emission mechanisms of GRBs cannot be attributed to the structured jet model.
% is ruled out for the emission mechanisms for
%the prompt emission of GRBs.
this difficulty can be overcome if the jet possesses a more complex structure.
For example, if the jet has more than two components and
velocity shear is present at multiple angles 
more closely spaced than $\Gamma^{-1}$,
photons from the acceleration regions will be prominent 
for all observers lying within the opening angle of the jet
($\theta_{\rm obs} \leq \theta_1$).
Even in the case of a simple spine-sheath jet,
if the angular boundary of the spine and sheath ($\theta_0$) varied rapidly with 
time, 
the accelerated non-thermal photons 
would be observable 
across a broad range of angles.
It should be also noted that the structure  
need not be in the direction of $\theta$.
Strong velocity shear in the azimuthal ($\phi$) direction and/or
radial direction due to the presence of  turbulence
or shocks can also provide an acceleration sites for the photons 
\citep{BML11,IOK11}.
Any structure showing strong velocity shear within an 
angle $\Gamma^{-1}$ 
from the line-of-sight can give rise
to a non-thermal component above the thermal peak.
Hence, it is expected that jet having a rich structure and/or
rapid time variability will be naturally accompanied by
non-thermal emission, irrespective of the observer angle.
The origin of the structure and variability could be due to 
the nature of the central engine
\citep{Mc06, NTM07, N09, MB09, N11}
 and/or the propagation of jet 
through the envelope of
 the progenitor star \citep{ZWM03, MYN06, MLB07, LMB09, MNA11, NIK11}.
%As for the latter case, indeed a
%number of numerical simulations have shown 
%that the jet will have rich structures within the flow
%due to interaction with the progenitor star
%in various spatial scales \citep[e.g.,][]{LMB09, MNA11, NIK11}.

%\subsection{On the Observer Angle Dependence and Structure of the Jet}

\subsection{On the $E_{\rm p}$-$L_{\rm p}$ relation}

Our calculations have shown that the peak energy
of the  observed spectra can be roughly approximated
as the thermal peak
 of a blackbody emission from the surface of optical depth
$\tau(r) \sim 2$ ($r \sim r_{\rm ph}/2$).
 On the other hand, 
 the peak  luminosity shows rough agreement with
 that of the emission from $\tau(r) \sim 1$ ($r \sim r_{\rm ph}$).
% $\tau(r) \sim 1$ ($r \sim r_{\rm ph}$), respectively.
%, while the peak  luminosity 
%roughly agrees with that for 
This result is valid for an adiabatic fireball which has a photosphere above the saturation radius
($r_{\rm ph} > r_{\rm s}$)
(see \S\ref{uniform} for detail).
%main part (thermal component) of the observed spectra 
%can be approximated as a 
%blackbody  from a surface of optical depth $\tau(r) \sim 5$ ($r \sim r_{\rm ph}/5$)
%for an adiabatic fireball which has a photosphere above the saturation radius
%($r_{\rm ph} > r_{\rm s}$)
As a result, the peak energy and luminosity are given by
\begin{eqnarray}
\label{Epeak}
E_{\rm p}\sim
 800 r_{\rm i,8}^{1/6}\eta_{400}^{8/3}L_{53}^{-5/12}~{\rm keV},
\end{eqnarray}
and 
\begin{eqnarray}
\label{Lpeak}
L_{\rm p} \sim  10^{52} r_{\rm i,8}^{2/3}\eta_{400}^{8/3}L_{53}^{1/3}~{\rm erg/s},
\end{eqnarray}
 respectively 
\footnote{For  cases where  efficient energy dissipation 
is present within the flow, the dependence of
 $E_{\rm p}$ and $L_{\rm p}$
on the fireball parameters can be significantly
different \citep[e.g.,][]{RM05, G12, LMM13}.
It is worthy to note that 
  comparison of Eqs. (\ref{Epeak}) and (\ref{Lpeak})
 with the observed peak energy and luminosity will 
enable us to constrain the properties (fireball parameters)
 of the GRB jet \citep[e.g.,][]{PRW07,FWZ12}.
}.
Derived from observations,
there is an empirical relation between the peak energy and
luminosity \citep{YMN04, KYM08, NGG11, AFT02, WG03} that is roughly given by
\begin{eqnarray}
\label{ELRel}
E_{\rm p} \approx 600 \left(\frac{L_{\rm p}}{10^{53}~{\rm erg/s}}\right)^{1/2}~{\rm keV} . 
\end{eqnarray}
Therefore, in order to reproduce the empirical relation
from the photospheric emission,
the parameters of the fireball ($r_{\rm i}$, $\eta$, and $L$) must satisfy 
\begin{eqnarray}
\label{const1}
%L \sim 10^{54} r_{\rm i,8}^{2/3}\eta_{400}^{8/3}~{\rm erg/s}.
L_{53} \sim 12 r_{\rm i,8}^{-2/7}\eta_{400}^{16/7}~{\rm erg/s}.
\end{eqnarray}

Another important ingredient for
comparison with observations is the 
emission efficiency 
$\eta_{\gamma}=
L_{\rm p}/L$. % \sim 0.35 r_{\rm i,8}^{2/3}\eta_{400}^{8/3}L_{53}^{-2/3}$. 
Observations of the afterglows suggest a quite high GRB efficiency
in the range $0.01 \lesssim \eta_{\gamma} \lesssim 1$,
with a typical value at $\eta_{\gamma}\sim 0.1-0.2$ 
\citep[e.g.,][]{FP06,ZLP07}. %
%
%Hence, to be consistent with the observations,
%efficiency must satisfy $0.1 \lesssim \eta_{\gamma} \leq 1$.
This gives  another constraint on the fireball parameters
which can be written as 
\begin{eqnarray}
\label{const2}
0.01 \lesssim   0.1 r_{\rm i,8}^{2/3}\eta_{400}^{8/3}L_{53}^{-2/3} \lesssim 1.
\end{eqnarray}
To sum up, 
the observed 
 $E_{\rm p}$-$L_{\rm p}$ relation
as well as the high efficiency %suggested by observations
can be reproduced  when
equations (\ref{const1}) and (\ref{const2}) are satisfied.

In Fig. \ref{EpLpfig},
we show the calculated spectra for  a uniform jet
($\theta_1 = \theta_0 = 1^{\circ}$) 
in which both conditions are fulfilled.
The adopted values of the parameters are summarized in the figure.
The red, green, blue and purple lines correspond to the cases of
$\eta_0 = 100$, $\eta_0 = 200$, $\eta_0 = 400$ and $\eta_0 = 500$,
respectively.
The remaining parameters ($L_0$ and $r_{\rm i}$) 
are determined
so that an efficiency of
$\eta_{\gamma} \sim 0.1 r_{\rm i,8}^{2/3}\eta_{400}^{8/3}L_{53}^{-2/3} 
\sim 0.2$ is realized. 
The thick light blue line displays the observed $E_{\rm p}$-$L_{\rm p}$
 relation.
From the figure,
it can be confirmed that the photospheric emission 
can indeed reproduce
the empirical
$E_{\rm p}$-$L_{\rm p}$ relation 
while at the same time retaining a high efficiency
when the two conditions are satisfied.

%
%{\bf COMMENTS on the employed fireball parameters required.}
%

%The efficiency of the emission is roughly given as
%$\epsilon = L_{\rm p}/L
%  \sim 3.5\times 10^{-2} r_{\rm i,8}^{-1/6}\eta_{400}^{8/21}L_{53}^{1/3}$.
%$\epsilon = L_{\rm p}/L
%  \sim 0.35 r_{\rm i,8}^{2/3}\eta_{400}^{8/3}L_{53}^{-2/3}$.
%
%$L \lesssim 6.5 \times 10^{53} \epsilon_{-1}^{-3/2}r_{\rm i, 8} \eta_{400}^4
%~{\rm erg/s}$

%The condition for $\epsilon \leq 1$ leads to
%$L \gtrsim 2.1 \times 10^{52} r_{\rm i, 8} \eta_{400}^4~{erg/s}$ 

\subsection{Dependence on the width of the spine-sheath boundary layer}

In the present study,
 we have assumed an infinitesimal width for the boundary layer of
the spine and sheath region.
However, in reality, the boundary layer is expected to possess a finite width
due to the interaction between the two regions.
Firstly, by definition, photons are closely coupled to 
the matter below the photosphere, and, therefore, 
% are traveling across the boundary layer
will couple the two regions due to  Compton friction (radiative viscosity).
In particular,
this effect will be important in the regions where the energy is dominated by 
radiation ($r \lesssim r_{\rm s}$) \citep[e.g., ][]{AB92}.
Secondly, even in the absence of radiation coupling,
the Kelvin-Helmholtz instability
should grow whenever velocity shear is present \citep [e.g.,][]{TS76, BMR04}.
These effects will relax the discontinuous change in the velocity
and lead to broadening of the boundary layer.
Although the detail analysis of  the resultant jet structure 
is beyond the scope of the present study,
these effects will reduce the acceleration of the photons.
To quantify the effect of the  broadening of the boundary layer on the
spectra, here
 we compute the photon propagation within a spine-sheath 
jet having a boundary layer with finite width %$d\theta_{\rm B}$
 in transverse direction.
We  explore
the dependence on the width of the boundary layer by considering 
cases with various widths.

In modeling the jet structure,
we added slight modifications in the original spine-sheath jet 
model considered in the present study.
The boundary layer is defined as a region having finite transverse width 
$d\theta_{\rm B}$ that is located in the range
% $\theta_0 - d\theta_{\rm B}/2 < \theta < \theta_0 + d\theta_{\rm B}/2$.
 $\theta_{\rm i} < \theta < \theta_{\rm ii}$, where
$\theta_{\rm i} = \theta_0 - d\theta_{\rm B}/2$ and $\theta_{\rm ii} = \theta_0 + d\theta_{\rm B}/2$.
%The tansverse structure of the jet is modeled 
%by  parameterizing the finite transverse width of the spine-sheath boundary layer
%as $d\theta_{\rm B}$.
Correspondingly, the  spine and sheath
 regions are limited to the ranges
% $0 \leq \theta \leq \theta_0 - d\theta_{\rm B} / 2$
 $0 \leq \theta \leq \theta_{\rm i}$
 and
% $\theta_0 + d\theta_{\rm B}/2 \leq \theta \leq \theta_1$,
  $\theta_{\rm ii} \leq \theta \leq \theta_1$,
 respectively.
While the fluid
 properties (fireball parameters) of the spine and sheath region
are determined in the same way 
as in the cases of infinitesimal boundary layer (\S\ref{SPst}),
the properties of the boundary layer are determined by simply
imposing a linear interpolation of the fireball parameters
from the two regions.
Hence, the 
 initial radius of the fireball is fixed  in all regions
 ($r_{\rm i}=10^{8}~{\rm cm}$), and
 the transverse distribution of the dimensionless entropy and kinetic
 luminosity within is given by
\begin{eqnarray}
\label{eta_dis}
\eta(\theta) =   \left\{ \begin{array}{lcrll}
      \eta_0 &~~
         {\rm for}~~ & 0 \leq & \theta & \leq \theta_{\rm i}  ,  \\
      \frac{(\theta - \theta_{\rm i})\eta_1 + (\theta_{\rm ii} - \theta)\eta_0}{d\theta_{\rm B}} &~~
         {\rm for}~~ & \theta_{\rm i} < & \theta & < \theta_{\rm ii} ,  \\
     \eta_1  &~~
         {\rm for}~~ & \theta_{\rm ii} \leq  & \theta & \leq \theta_1  , \\
             \end{array} \right. 
\end{eqnarray}
and
\begin{eqnarray}
\label{kin_dis}
L(\theta) =   \left\{ \begin{array}{lcrll}
      L_0 &~~
         {\rm for}~~ & 0 \leq & \theta & \leq \theta_{\rm i}  ,  \\
      \frac{(\theta - \theta_{\rm i})L_1 + (\theta_{\rm ii} - \theta)L_0}{d\theta_{\rm B}} &~~
         {\rm for}~~ & \theta_{\rm i} < & \theta & < \theta_{\rm ii} ,  \\
     L_1  &~~
         {\rm for}~~ & \theta_{\rm ii} \leq  & \theta & \leq \theta_1  , \\
             \end{array} \right. 
\end{eqnarray}
respectively. 
Hence,
the  profile of the velocity (Lorentz factor) and density
within the jet is continuous 
in all regions.

Having determined the background fluid properties,
the propagation of photons is calculated in the same way as in the case
of an infinitesimal boundary layer.
Black body emission 
is injected at the surface of a fixed radius $r_{\rm in} = r_{\rm s1}$
with a comoving temperature determined from the fireball parameters 
(\S\ref{Iniph}).
Then, photons are propagated until they reach the outer boundary 
of the calculation range  (see \S\ref{Phtr} for detail).

In Figs. \ref{G42sm}
and \ref{G41sm}, we show  the
 obtained  spectra for
a stratified jet in which the fireball parameters for the spine and sheath
regions  are identical to those employed in 
Case I ($\eta_0 = 400$ and $\eta_1 = 200$)
 and Case II  ($\eta_0 = 400$ and $\eta_1 = 100$)
discussed in \S\ref{crdiss}.
% $\eta_0=400$ and $L_0=10^{53}~{\rm erg/s}$ for
%the spine, and 
%$\eta_1=200$ and  $L_1= (\eta_1/\eta_0)L_0 = 5\times10^{52}~{\rm erg/s}$
%for the sheath.
The various lines reflect the width of the boundary layer
$d\theta_{\rm B}$.
While the thick red line displays the case for the infinitesimal width
 ($d\theta_{\rm B}=0$),
the thin green, blue, purple, light blue, yellow, black and red lines 
display the cases for the finite widths of
 $d\theta_{\rm B}=(100\eta_0)^{-1}$,  
 $(20\eta_0)^{-1}$,  
 $(10\eta_0)^{-1}$,  
 $(5\eta_0)^{-1}$,  
 $(2\eta_0)^{-1}$,  
 $\eta_0^{-1}$
and  $(\eta_0/2)^{-1}$, respectively.
In all cases, the observer angle is fixed at
 $\theta_{\rm obs}=\theta_0 = 0.5^{\circ}$.
As expected,
%From the figure, it is seen that
the non-thermal component becomes softer
as the width of the boundary layer broadens
due to the reduction in efficiency of photon acceleration.
Note that, 
while the broadening of the boundary layer causes the 
spectra to depart from the
high energy tail of the  Band function for Case I, 
good agreement is found at $d\theta_{\rm B}\sim (5\eta_0)^{-1}$ for Case II.
Therefore, although a large gradient in the Lorentz factor is required,
we emphasize that the broadening of the 
boundary layer does not rule out the photon acceleration mechanism in 
a stratified jet as the origin of the
high energy spectra in the prompt emission.

From the figures, it is seen that, in order to provide an
efficient acceleration site comparable to the case of 
boundary layer with infinitesimal with,
the bulk Lorentz factor must vary by a factor of few
roughly within an angle  
$d\theta_{B} \sim (100\eta_0)^{-1} \ll \Gamma^{-1}$.
As is expected, 
this condition is roughly
equivalent to the condition for the
boundary layer to be optically thin 
above the  radius where the velocity shear begins to develop
 ($r\geq r_{\rm s1}$).
% the 
%photons to cross the boundary layer without being scattered
%above the radius where the velocity shear develops ($r\geq r_{\rm s1}$).
This can be shown as follows.
The typical angle between the photon propagation direction and the 
 fluid velocity (radial) direction is roughly
$\langle{\theta_v}\rangle \sim \Gamma^{-1}$.
Hence,
 at a
given radius $r$,
the typical distance that the photons must propagate
 to cross the boundary layer is roughly given as 
 $l_{\rm cr} \sim r (d\theta_{\rm B}/\langle{\theta_v}\rangle) 
\sim r \Gamma d\theta_{\rm B}$.
The above estimate is valid as long as $d\theta_{\rm B} \ll \Gamma^{-1}$.
On the other hand, the mean free path
for the typical photon at a given radius $r$ is
roughly given by
$l_{\rm mfp} \sim
  {\cal D}(\langle \theta_v \rangle)^{-1}\sigma_{\rm T}n_e(r) \sim
 r / \tau(r)$.
Thus, the typical optical depth for a photons to cross a
boundary layer of  finite width can be 
estimated as
$\tau_{\rm cr} \sim 
l_{\rm cr} /   l_{\rm mfp}
\sim \tau(r) \Gamma d\theta_{\rm B}$.
Since $\tau(r_{\rm s1})$ and $\Gamma(r_{\rm s1})$ 
are $\sim 100$  ($\sim 600$) and $1/2 \eta_0$ ($1/4 \eta_0$)
for Case I (Case II),  respectively,
$\tau_{\rm cr} \lesssim 1$ is obtained at $r=r_{\rm s1}$
for $d\theta_{\rm B} \lesssim (100 \eta_0)^{-1}$. 
As a result,
since  $\tau_{\rm cr}$ decreases as the 
radius increases, when 
the condition $d\theta_{\rm B} \lesssim (100 \eta_0)^{-1}$
is satisfied, the boundary layer is optically thin to
crossings at $r\geq r_{\rm s1}$.

 Lastly, 
 let us briefly comment on a %comparison with the
 similar calculation performed by \citet{LPR13}.
% While we consider a photon propagation in a
% two-component jet having discontinuity at $\theta = \theta_0$,
 In their study,
 photon propagation within a
 jet having a continuously  decaying velocity profile
 $\Gamma \propto \theta^{-p}$
 at the  outer region ($\theta \gtrsim \theta_{\rm j}$) is explored. 
 Since the efficiency of  photon acceleration 
 increases as the 
 difference in velocity (Lorentz factor)
 between the spine and sheath region increases,
% within a small
% angular scale ($\theta \lesssim \Gamma^{-1}$) increases,
 their calculation should also show
 non-thermal high energy photons
 when a large velocity gradient is considered. 
% Hence, 
% their
% calculation should also show acceleration of photons 
% when large velocity gradient is considered. 
 Indeed, although not so prominent as is shown by our results,
signs of photon acceleration are also reported  in their study.
 While only a thermal component is present when a
 relatively small velocity gradient ($p=2$)
 is  assumed,
 power-law excess %($\beta_{\rm ph} \ sim -2$) 
 above the thermal peak energy appears in the calculated
 spectra
 when a large velocity gradient ($p=4$) is assumed.
 Therefore, as the velocity gradient enlarges,
 we expect to find more efficient acceleration as
 is seen in the present study.

\subsection{On the Absorption Processes}
Since the photon number density is
significant  inside the jet
($r\lesssim r_{\rm ph}$), 
photons are subject to 
 $\gamma \gamma$ attenuation
once they exceed the threshold energy for the process.
The threshold energy is given by
$E_{\rm th}= 2 (m_e c^2)^2 / [(1-{\rm cos}\theta_{\rm ph})\epsilon]$,
where $\epsilon$ and $\theta_{\rm ph}$ are
the energy of the target photons and angle between the
propagation direction of the target and incident photons, respectively.
Since the photons are basically advected in the radial direction
within a small angle $\Gamma^{-1}$,
the collision angle of the photons is expected to be in a range
$\theta_{\rm ph} \lesssim \Gamma^{-1}$.
Hence, for photons confined in an angle  $\Gamma^{-1}$,
the minimum value of the threshold energy 
($\theta_{\rm ph} \sim \Gamma^{-1}$)
can be roughly estimated as
$E_{\rm th, min} \sim 100~(\Gamma / 100)^2
(\epsilon / 100~{\rm MeV})^{-1}~{\rm MeV}$.
For the cases considered in the present study 
($\Gamma \geq 100$), the photon energies do not 
exceed  $100~{\rm MeV}$ by much and, therefore,
the pair creation process is negligible for the majority of the photons.
However,
 for the small fraction of photons which propagate with a large angle
($\gtrsim \Gamma^{-1}$) 
with respect to the radial direction, this effect can become non-negligible.
In particular,
pair annihilation may play an important role for the high energy photons  which
are being accelerated, since 
scattering with a large angle with respect to the fluid velocity
(radial) direction is favored for gaining photon energy (see \S\ref{model} for detail).
To check this, we have compared our results 
with those obtained by discarding the photons which
have exceeded the threshold energy 
obtained by substituting 
 the angle between the 
photon propagation direction and radial direction 
for $\theta_{\rm ph}$ and 
the highest photon energy appear in the calculation $\sim \Gamma_0 m_e c^2$
for $\epsilon$.
%(Since highest energy is considered as the targe photon energy,
% this can safely determine the upper limit on the significance of 
%parir annihilation)
% highest target photon energy $\epsilon = \Gamma_0 m_e c^2$
%before reaching the outer boundary of the calculation.
Indeed, we find that only a small fraction of photons are absorbed
and the change in the spectrum is negligible.
Therefore, we conclude that $\gamma \gamma$ attenuation effect does
not affect the obtained results.

The flow within  the fireball is expected to be fully ionized,
and the effect of free-free absorption should also be discussed.
For an electron-proton plasma, the
frequency averaged free-free opacity can be roughly written as 
$\alpha_{\rm ff}(r) \approx 1.7\times 10^{-25} T'(r)^{-7/2}n_e(r)^2~{\rm cm}^{-1}$ \citep{RL79} in the fluid rest frame.
In the laboratory frame, by using a Doppler factor,
 the opacity can be  expressed as
$\alpha_{\rm ff, lab}={\cal D}^{-1}\alpha_{\rm ff}$.  % \sim \Gamma^{^1}\alpha_{ff}$.
Since photons advect in the radial direction within an angle
 $\Gamma^{-1}$ (${\cal D} \sim \Gamma$),
the optical depth for photons which have propagated from
$r = r_{\rm inj}$ to the observer (infinity) can be roughly estimated as
$\tau_{\rm ff} \sim r_{\rm inj}\alpha_{\rm ff, lab}\Gamma^{-1}\sim
 8.3\times 10^{-6} (L/10^{53}~{\rm erg~s^{-1}})^{9/8} (\eta/400)^{-2}
 (r_{\rm i}/10^8~{\rm cm})^{5/4}$ $(r_{\rm inj}/10^{10}~{\rm cm})^{-5/2}$, %--- maybe make this into an inline equation?} where
we have assumed $r_{\rm inj}\leq r_{\rm s}$
and used equations (\ref{ne}) and (\ref{Tev}) in the last equality.
Therefore, we can conclude that free-free absorption is 
also negligible.

\subsection{On the GeV Gamma-ray}
\label{GeV}

Fermi observations have shown that
a fraction ($\sim 8 \%$) of GRBs are accompanied by
significant emissions at energies well above $\sim 100{\rm MeV}$
 \citep[e.g.,][]{ZZ11,Fermi12}.
Within the framework of our model,
the energy of the photons is limited by the bulk Lorentz factor
of the flow as $h \nu_{\rm obs} \lesssim 50(\Gamma / 100)~{\rm MeV}$ (\S\ref{result}).
Hence, in order to generate  emissions at $h \nu_{\rm obs} \sim {\rm GeV}$,
large bulk Lorentz factors such as  $\Gamma > 2000$ are required.
Alternatively,
if we consider the presence of relativistic electrons due to
some kind of dissipative processes within the flow
\citep[e.g.,][]{IMT07, LB10,G12, B10, VBP11}, 
lower values 
are allowed for the bulk Lorentz factor.
By denoting the maximum Lorentz factor of the electrons measured in
the rest frame of the fluid as $\gamma_{\rm max}$, 
the lower limit on the bulk Lorentz factor to produce ${\rm GeV}$ photons
decreases as $\Gamma > 200 (\gamma_{\rm max}/10)^{-1}$.
Therefore, within this picture,
GRBs with intense ${\rm GeV}$ emissions may imply the
presence of
fluid components with very  high bulk Lorentz factors ($\Gamma > 2000$)
or a dissipative process producing relativistic electrons
within the flow.
In either case, pair cascades due to
$\gamma \gamma$ attenuation may play an important role
\citep{IOK11},
and the details of the process and its effect on the observed
photon spectra
are beyond the scope of the present study.

It should be also noted that
photons with energies above and below
$\sim 100~{\rm MeV}$ may have distinct origins.
For example,
it is widely discussed that 
$\sim {\rm GeV}$ emission can result from energy dissipation by external
shocks similar to afterglow emission \citep{KD09,KD10,GGN10},
due to the smooth temporal decay
seen in lightcurve of many LAT detected GRBs.
Therefore, one possible interpretation is that,
while emission at $\lesssim 100~{\rm MeV}$ has a photospheric origin
as discussed in the present study,
higher energy photons are independently produced at 
the external shock.
%
%Though very 
%
%In this case, 
%either high $\Gamma$ or $\gamma_{\rm max}$
%are not necessary
%(though independent constraint on flow parameters
%can be obtained from the external shock model).

\section{SUMMARY AND CONCLUSIONS}
\label{summary}
In the present study, we have explored  photospheric emission from 
ultra-relativistic jets which have a velocity shear in the transverse
direction.
For the jet structure, we considered  a two component outflow 
in which a fast spine jet is embedded in a slower sheath region.
The fluid properties such as electron number density $n_{\rm e}(r)$
and bulk Lorentz factor $\Gamma(r)$ 
are determined by applying the
adiabatic fireball model %which is characterized
in each region independently.
Initially thermal photons are injected at a radius far below the
photosphere ($r_{\rm inj}=\eta_1 r_{\rm i} \ll r_{\rm ph}$)
in which the velocity shear begins to develop. 
Using a Monte-Carlo technique, 
injected photons propagate
until they reach the outer boundary, located at a radius
where the optical depth is small ($\tau \ll 1$).
The following is a summary of 
the main results and conclusions of the present study.

\noindent{1.} 
Due to the presence of velocity shear,
photons that cross the boundary between the  spine and sheath ($\theta_0$)
multiple times
can gain energy through a Fermi-like acceleration mechanism.
The acceleration process 
at the boundary layer can proceed efficiently until
the photon reaches an energy where the Klein-Nishina effect becomes important.
As a result, the maximum energy of the accelerated photons is 
limited by
the bulk Lorentz factor of the outflow as
$\sim \Gamma_0 m_e c^2 \sim 200(\Gamma_0/400)~{\rm MeV}$. 
These accelerated photons produce a
non-thermal component above the thermal peak
in the observed spectra.
Although it depends strongly on the flow profile,
the non-thermal component can reproduce the
high energy spectra of typical GRBs 
 ($\nu L_{\nu} \propto \nu^{-0.5}$).
The accelerated photons
may also be capable of 
explaining the extra hard power-law component 
above the bump of the thermal-like peak seen in some
peculiar bursts (GRB 090510, 090902B, 090926A).

%acceleration processes proceed efficiently 
%until the photon reach en

%owning to the Klein-Nishina effect.
%
%These accelerated photons produces a
%non-thermal component above the thermal peak
%in the observed spectra.
%The maximum energy of the photons is 
%limited by
%the bulk Lorentz factor of the outflow as
%$\sim \Gamma_0 m_e c^2 \sim 200(\Gamma_0/400)~{\rm MeV}$. 
%owning to the Klein-Nishina effect.

\noindent{2.}
The efficiency of the acceleration is sensitive to 
the relative difference of the bulk Lorentz factor
in the two regions ($\Gamma_0/\Gamma_1$),
the optical depth
at the radius where the velocity shear begins to develop
($\tau(r_{\rm inj})$)
and the optical depth for a photon to cross the 
boundary transition layer of the two regions ($\tau_{\rm cr}$).
For an efficient acceleration, 
larger values are favored for  $\Gamma_0/\Gamma_1$ and $\tau(r_{\rm inj})$,
while smaller value is favored for $\tau_{\rm cr}$,
%Larger values of $\Gamma_0/\Gamma_1$ and $\tau(r_{\rm inj})$
%are favored for efficient acceleration,
since both the
energy gain per crossing and probability for the photons to cross
the boundary layer increase.
The increase in the efficiency 
leads to a  harder high energy non-thermal component in
the observed spectra.

\noindent{3.}
The observed
spectra strongly depend on the observer angle $\theta_{\rm obs}$
due to the relativistic beaming effect.
The high energy non-thermal component is hardest when the 
observer is aligned to the boundary layer ($\theta_{\rm obs} = \theta_0$)
and becomes softer as the difference between $\theta_{\rm obs}$ and $\theta_0$
becomes larger.
Additionally, the non-thermal component is most prominent
for an observer located near the boundary layer
($|\theta_0 - \theta_{\rm obs}| \lesssim \Gamma^{-1}$), and
it becomes significantly weaker or absent
when the observer angle is far from the boundary layer
($|\theta_0 - \theta_{\rm obs}| \gtrsim \Gamma^{-1}$).
In order for the intense non-thermal component 
to be seen for all observers in a range 
$\theta_{\rm obs} \lesssim \theta_1$,
a multi-component jet in which velocity shears
are present in an
interval of angles smaller than $\Gamma^{-1}$ is required.

\noindent{4.}
The observed spectra below the peak energy 
are determined by the majority of
thermal photons which have not experienced acceleration.
The spectra $\nu L_\nu \propto \nu^{2.4}$ are somewhat softer than
that expected from the Rayleigh-Jeans part of a
single-temperature blackbody emission
($\nu L_{\nu} \propto \nu^3$), since
the emission is a superposition of 
photons released at various angles 
which have different observed temperatures.
This is still much harder than the typical
low energy spectral index of the observed GRBs ($\nu L_\nu \propto \nu$),
implying that  the steady outflow component has difficulty in
reproducing the overall spectra.
Hence, time evolution of outflow properties seems to be required.
It is indeed shown that time-integrated spectra
of an unsteady outflow
 can reproduce
the low energy spectra due to the multi-temperature effect.

\noindent{5.}
Photons begin to decouple from the matter below the photosphere
typically at $r \sim r_{\rm ph}/5$,
irrespective of the imposed fireball parameters
($r_{\rm in}$, $L$ and $\eta$) in the background fluid.
The resultant peak energy $E_{\rm p}$ and luminosity $L_{\rm p}$
can be roughly approximated by the corresponding values
of a blackbody emission from the surface of
 $r\sim r_{\rm ph}/2$ and
 $r\sim r_{\rm ph}$, respectively.
The empirical $E_{\rm p}$-$L_{\rm p}$
relation can be well reproduced
by considering   the difference in the outflow properties 
of individual sources.

%if multi-temperature effect due to time evolution of outflow properties are considered.
%
%Alternatively, the  multi-temperature effect due to 
%the continuous change of the flow properties in
%lateral direction is also a possible origin for the 
%low energy slope of the observed spectra
%as discussed in \citet{LPR12}.

%\ni

%As noted previously,
%we solve the scattering between the photon due
%to the electrons in this simulation.
%For simplicity, we negelect the thermal motion of the electrons
%and assume that the electrons have zero temperature.
%
%The calculation 
%Under the above assumptions, full Klein-Nishina cross section
%is used in calculating the propagation of photons.

%In GRB jets, opacity of photons is strongly dominated by the scatterings 
%due to electrons.
%Hence, we evaluate the transfer of photons 
%by only considering the scattering process.

%\begin{eqnarray}
%T^{\mu \nu} = \rho_{0} h u^{\mu} u^{\nu} + p g^{\mu \nu}
%\end{eqnarray}

%\citet{}            xxx et al. (xxxx)
%\citep{}            (xxx et al xxxx)
%\citep[bla][]{}     (bla xxx et al xxxx)

%%%%%%%%%%%%%%%%%%%%%%%%%%%%%%%%%%%%%%%%%%%%%%%%%%%%%%%%%%%

\acknowledgments 

We thank Don Warren for comments and assistance that improved
the clarity of the paper.
We also thank Alexey Tolstov, Yudai Suwa and Yuichiro Sekiguchi for
fruitful discussions.
We are grateful to the anonymous referee for constructive comments.
This work is supported by the support from
the Ministry of Education,
Culture, Sports, Science and Technology
(No. 20105005, No. 23105709 and No. 24244036), the
Japan Society for the Promotion of Science (No. 19104006 and
No. 23340069), and the Global COE Program, ``The Next Generation
of Physics, Spun from University and Emergence from MEXT of Japan''.
S.-H. L. and J. M. acknowledge 
support from Grants-in-Aid for Foreign JSPS Fellow (Nos. 2503018 and 24.02022).
We thank RIKEN for providing the facilities and financial support.


\begin{thebibliography}{}

%%%%%%%%%%%%%%%%%%%%%%%%%%%%%%%%%%%%%%%%%%%%%%%%%%%%%%%%%%%





%090902B
\bibitem[Abdo et al.(2009)]{AAA09} Abdo, A.~A., Ackermann, 
M., Ajello, M., et al.\ 2009, \apjl, 706, L138 

%090926A
\bibitem[Ackermann et al.(2011)]{AAA11} Ackermann, M., 
Ajello, M., Asano, K., et al.\ 2011, \apj, 729, 114 

%090510
\bibitem[Ackermann et al.(2010)]{AAA10} Ackermann, M., Asano, 
K., Atwood, W.~B., et al.\ 2010, \apj, 716, 1178 



\bibitem[Amati et al.(2002)]{AFT02} Amati, L., Frontera, F., Tavani, M., et al.\ 2002, \aap, 390, 81 


\bibitem[Arav \& Begelman(1992)]{AB92} Arav, N., \& Begelman, M.~C.\ 1992, \apj, 401, 125 



\bibitem[Band et al.(1993)]{BMF93} Band, D., Matteson, J., 
Ford, L., et al.\ 1993, \apj, 413, 281 


\bibitem[B{\'e}gu{\'e} et al.(2013)]{BSV13} B{\'e}gu{\'e}, 
D., Siutsou, I.~A., \& Vereshchagin, G.~V.\ 2013, \apj, 767, 139 


\bibitem[Beloborodov(2010)]{B10} Beloborodov, A.~M.\ 2010, 
\mnras, 407, 1033 


\bibitem[Beloborodov(2011)]{B11} Beloborodov, A.~M.\ 2011, 
\apj, 737, 68 


\bibitem[Beloborodov(2013)]{B13} Beloborodov, A.~M.\ 2013, 
\apj, 764, 157 




\bibitem[Bodo et al.(2004)]{BMR04} Bodo, G., Mignone, A., 
\& Rosner, R.\ 2004, \pre, 70, 036304 


\bibitem[Bromberg et al.(2011)]{BML11} Bromberg, O., 
Mikolitzky, Z., \& Levinson, A.\ 2011, \apj, 733, 85 


\bibitem[Budnik et al.(2010)]{BKS10} Budnik, R., Katz, B., 
Sagiv, A., \& Waxman, E.\ 2010, \apj, 725, 63 



\bibitem[Crider et al.(1997)]{CLS97} Crider, A., Liang, 
E.~P., Smith, I.~A., et al.\ 1997, \apjl, 479, L39 


\bibitem[Eichler(1994)]{E94} Eichler, D.\ 1994, \apjs, 90, 
877 


% Compact fireball 
\bibitem[Eichler \& Levinson(2000)]{EL00} Eichler, D., \& Levinson, A.\ 2000, \apj, 529, 146 


\bibitem[Fan \& Piran(2006)]{FP06} Fan, Y., \& Piran, T.\ 2006, \mnras, 369, 197 
\bibitem[Fan et al.(2012)]{FWZ12} Fan, Y.-Z., Wei, D.-M., 
Zhang, F.-W., \& Zhang, B.-B.\ 2012, \apjl, 755, L6 


\bibitem[Ghirlanda et al.(2003)]{GCG03} Ghirlanda, G., Celotti, A., \& Ghisellini, G.\ 2003, \aap, 406, 879 


\bibitem[Ghisellini et al.(2010)]{GGN10} Ghisellini, G., 
Ghirlanda, G., Nava, L., \& Celotti, A.\ 2010, \mnras, 403, 926 

% magnetic recconection model
\bibitem[Giannios 
\& Spruit(2007)]{GS07} Giannios, D., \& Spruit, H.~C.\ 2007, \aap, 469, 1 

\bibitem[Giannios(2006)]{G06} Giannios, D.\ 2006, \aap, 457, 763 
\bibitem[Giannios(2008)]{G08} Giannios, D.\ 2008, \aap, 480, 305 

\bibitem[Giannios(2012)]{G12} Giannios, D.\ 2012, \mnras, 
422, 3092 

\bibitem[Goodman(1986)]{G86} Goodman, J.\ 1986, \apjl, 308, L47 

\bibitem[Guetta et al.(2001)]{GSW01} Guetta, D., Spada, M., 
\& Waxman, E.\ 2001, \apj, 557, 399 


\bibitem[Ioka et al.(2007)]{IMT07} Ioka, K., Murase, K., 
Toma, K., Nagataki, S., \& Nakamura, T.\ 2007, \apjl, 670, L77 




\bibitem[Ioka et al.(2011)]{IOK11} Ioka, K., Ohira, Y., 
Kawanaka, N., \& Mizuta, A.\ 2011, Progress of Theoretical Physics, 126, 555 


\bibitem[Jokipii et al.(1989)]{JKM89} Jokipii, J.~R., Kota, 
J., \& Morfill, G.\ 1989, \apjl, 345, L67 


\bibitem[Kaneko et al.(2008)]{KGP08} Kaneko, Y., 
Gonz{\'a}lez, M.~M., Preece, R.~D., Dingus, B.~L., 
\& Briggs, M.~S.\ 2008, \apj, 677, 1168 


\bibitem[Kaneko et al.(2006)]{KPB06} Kaneko, Y., Preece, 
R.~D., Briggs, M.~S., et al.\ 2006, \apjs, 166, 298 


\bibitem[Katz et al.(2010)]{KBW10} Katz, B., Budnik, R., 
\& Waxman, E.\ 2010, \apj, 716, 781 



\bibitem[Kino et al.(2004)]{KMY04} Kino, M., Mizuta, A., 
\& Yamada, S.\ 2004, \apj, 611, 1021 

\bibitem[Kobayashi et al.(1997)]{KPS97} Kobayashi, S., Piran, 
T., \& Sari, R.\ 1997, \apj, 490, 92 

% Yonetoku relation
\bibitem[Kodama et al.(2008)]{KYM08} Kodama, Y., Yonetoku, 
D., Murakami, T., et al.\ 2008, \mnras, 391, L1 


\bibitem[Kumar 
\& Barniol Duran(2009)]{KD09} Kumar, P., \& Barniol Duran, R.\ 2009, \mnras, 400, L75 


\bibitem[Kumar 
\& Barniol Duran(2010)]{KD10} Kumar, P., \& Barniol Duran, R.\ 2010, \mnras, 409, 226 



\bibitem[Lazzati \& Begelman(2010)]{LB10} Lazzati, D., \& Begelman, M.~C.\ 2010, \apj, 725, 1137 


\bibitem[Lazzati et al.(1999)]{LGC99} Lazzati, D., 
Ghisellini, G., \& Celotti, A.\ 1999, \mnras, 309, L13 

\bibitem[Lazzati et al.(2009)]{LMB09} Lazzati, D., Morsony, 
B.~J., \& Begelman, M.~C.\ 2009, \apjl, 700, L47 

\bibitem[Lazzati et al.(2013)]{LMM13} Lazzati, D., Morsony, 
B.~J., Margutti, R., \& Begelman, M.~C.\ 2013, arXiv:1301.3920 





\bibitem[Levinson(2012)]{L12} Levinson, A.\ 2012, \apj, 
756, 174 



\bibitem[Levinson 
\& Bromberg(2008)]{LB08} Levinson, A., \& Bromberg, O.\ 2008, Physical Review Letters, 100, 131101 



\bibitem[Levinson 
\& Globus(2013)]{LG13} Levinson, A., \& Globus, N.\ 2013, \apj, 770, 159 





\bibitem[Lundman et al.(2013)]{LPR13} Lundman, C., Pe'er, A., 
\& Ryde, F.\ 2013, \mnras, 428, 2430 
%\bibitem[Lundman et al.(2012)]{LPR12} Lundman, C., Pe'er, A., 
%\& Ryde, F.\ 2012, arXiv:1208.2965 
% theory of photospheric emission



\bibitem[McKinney(2006)]{Mc06} McKinney, J.~C.\ 2006, 
\mnras, 368, 1561 

\bibitem[McKinney 
\& Blandford(2009)]{MB09} McKinney, J.~C., \& Blandford, R.~D.\ 2009, \mnras, 394, L126 


\bibitem[M{\'e}sz{\'a}ros(2006)]{M06} M{\'e}sz{\'a}ros, P.\ 
2006, Reports on Progress in Physics, 69, 2259 


% internal shock photosphere for steep low photosphere required
\bibitem[M{\'e}sz{\'a}ros \& Rees(2000)]{MR00} M{\'e}sz{\'a}ros, P., \& Rees, M.~J.\ 2000, \apj, 530, 292 


\bibitem[Mizuta et al.(2011)]{MNA11} Mizuta, A., Nagataki, 
S., \& Aoi, J.\ 2011, \apj, 732, 26 

\bibitem[Mizuta et al.(2006)]{MYN06} Mizuta, A., Yamasaki, 
T., Nagataki, S., \& Mineshige, S.\ 2006, \apj, 651, 960 


\bibitem[Morsony et al.(2007)]{MLB07} Morsony, B.~J., 
Lazzati, D., \& Begelman, M.~C.\ 2007, \apj, 665, 569 

\bibitem[Nagakura et al.(2011)]{NIK11} Nagakura, H., Ito, H., 
Kiuchi, K., \& Yamada, S.\ 2011, \apj, 731, 80 


% central
\bibitem[Nagataki(2009)]{N09} Nagataki, S.\ 2009, \apj, 
704, 937 

\bibitem[Nagataki(2011)]{N11} Nagataki, S.\ 2011, \pasj, 
63, 1243 

\bibitem[Nagataki et al.(2007)]{NTM07} Nagataki, S., 
Takahashi, R., Mizuta, A., \& Takiwaki, T.\ 2007, \apj, 659, 512 

\bibitem[Nava et al.(2011)]{NGG11} Nava, L., Ghirlanda, G., 
Ghisellini, G., \& Celotti, A.\ 2011, \mnras, 415, 3153 


\bibitem[Ostrowski(1990)]{O90} Ostrowski, M.\ 1990, \aap, 238, 435 
\bibitem[Ostrowski(1998)]{O98} Ostrowski, M.\ 1998, \aap, 335, 134 




\bibitem[Paczynski(1986)]{P86} Paczynski, B.\ 1986, \apjl, 
308, L43 


% connection non-thermal + thermal  090902B
% thermal - non-thermal
\bibitem[Pe'er(2008)]{P08} Pe'er, A.\ 2008, \apj, 682, 463 

\bibitem[Pe'er et al.(2005)]{PMR05} Pe'er, A., 
M{\'e}sz{\'a}ros, P., \& Rees, M.~J.\ 2005, \apj, 635, 476 


\bibitem[Pe'er et al.(2006)]{PMR06} Pe'er, A., 
M{\'e}sz{\'a}ros, P., \& Rees, M.~J.\ 2006, \apj, 642, 995 



% Theory of multi-color BB
\bibitem[Pe'er \& Ryde(2011)]{PR11} Pe'er, A., \& Ryde, F.\ 2011, \apj, 732, 49 

% thermal fit to constrain fireball parameters
\bibitem[Pe'er et al.(2007)]{PRW07} Pe'er, A., Ryde, F., 
Wijers, R.~A.~M.~J., M{\'e}sz{\'a}ros, P., 
\& Rees, M.~J.\ 2007, \apjl, 664, L1 



\bibitem[Pe'er et al.(2012)]{PZR12} Pe'er, A., Zhang, B.-B., 
Ryde, F., et al.\ 2012, \mnras, 420, 468 


%review of GRB
\bibitem[Piran(2004)]{P04} Piran, T.\ 2004, Reviews of 
Modern Physics, 76, 1143 




% death line
\bibitem[Preece et al.(1998)]{PBM98} Preece, R.~D., Briggs, 
M.~S., Mallozzi, R.~S., et al.\ 1998, \apjl, 506, L23 


\bibitem[Preece et al.(2000)]{PBM00} Preece, R.~D., Briggs, 
M.~S., Mallozzi, R.~S., et al.\ 2000, \apjs, 126, 19 



\bibitem[Rees \& Meszaros(1994)]{RM94} Rees, M.~J., \& Meszaros, P.\ 1994, \apjl, 430, L93 


% dissipative photosphere model
\bibitem[Rees \& M{\'e}sz{\'a}ros(2005)]{RM05} Rees, M.~J., \& M{\'e}sz{\'a}ros, P.\ 2005, \apj, 628, 847 


% time-evo photosphere
\bibitem[Ruffini et al.(2011)]{RSV11} Ruffini, R., Siutsou, 
I.~A., \& Vereshchagin, G.~V.\ 2011, arXiv:1110.0407 


\bibitem[Rybicki 
\& Lightman(1979)]{RL79} Rybicki, G.~B., \& Lightman, A.~P.\ 1979, New York, Wiley-Interscience, 1979.~393 p.,  



% thermal + non-thermal GRB fit
\bibitem[Ryde(2005)]{R05} Ryde, F.\ 2005, \apjl, 625, L95 


\bibitem[Ryde et al.(2010)]{RAZ10} Ryde, F., Axelsson, M., 
Zhang, B.~B., et al.\ 2010, \apjl, 709, L172 

\bibitem[Ryde et al.(2006)]{RBK06} Ryde, F., Bj{\"o}rnsson, 
C.-I., Kaneko, Y., et al.\ 2006, \apj, 652, 1400 


\bibitem[Ryde et al.(2011)]{RPN11} Ryde, F., Pe'er, A., 
Nymark, T., et al.\ 2011, \mnras, 415, 3693 

\bibitem[Sari  \& Piran(1997)]{SP97} Sari, R., \& Piran, T.\ 1997, \apj, 485, 270 

%{\bf
%\bibitem[Suzuki \& Shigeyama(2013)]{SS13} Suzuki, A., \& Shigeyama, T.\ 2013, \apjl, 764, L12 
%}

\bibitem[The Fermi Large Area Telescope Team et 
al.(2012)]{Fermi12} The Fermi Large Area Telescope Team, 
Ackermann, M., Ajello, M., et al.\ 2012, \apj, 754, 121 


% large baryon load fireball photosphere + boost thermal photons by scattering of MHD turburlence
\bibitem[Thompson(1994)]{T94} Thompson, C.\ 1994, \mnras, 
270, 480 


\bibitem[Turland \& Scheuer(1976)]{TS76} Turland, B.~D., \& Scheuer, P.~A.~G.\ 176, \mnras, 176, 421 


\bibitem[Vurm et al.(2011)]{VBP11} Vurm, I., Beloborodov, 
A.~M., \& Poutanen, J.\ 2011, \apj, 738, 77 


\bibitem[Vurm et al.(2013)]{VLP13} Vurm, I., Lyubarsky, Y., 
\& Piran, T.\ 2013, \apj, 764, 143 



\bibitem[Wei 
\& Gao(2003)]{WG03} Wei, D.~M., \& Gao, W.~H.\ 2003, \mnras, 345, 743 




%\bibitem[Xu et al.(2011)]{XNH11} Xu, M., Nagataki, S., 
%\& Huang, Y.~F.\ 2011, \apj, 735, 3 

\bibitem[Xu et al.(2012)]{XNH12} Xu, M., Nagataki, S., Huang, 
Y.~F., \& Lee, S.-H.\ 2012, \apj, 746, 49 


\bibitem[Yonetoku et al.(2004)]{YMN04} Yonetoku, D., 
Murakami, T., Nakamura, T., et al.\ 2004, \apj, 609, 935 



% Comprehensive FERMI GRB
\bibitem[Zhang et al.(2011)]{ZZ11} Zhang, B.-B., Zhang, B., 
Liang, E.-W., et al.\ 2011, \apj, 730, 141 


\bibitem[Zhang et al.(2007)]{ZLP07} Zhang, B., Liang, E., 
Page, K.~L., et al.\ 2007, \apj, 655, 989 



\bibitem[Zhang et al.(2003)]{ZWM03} Zhang, W., Woosley, 
S.~E., \& MacFadyen, A.~I.\ 2003, \apj, 586, 356 






















\end{thebibliography}
\end{document}